\documentclass[fleqn,usenatbib]{mnras}

\usepackage{newtxtext,newtxmath}

\usepackage[T1]{fontenc}

\DeclareRobustCommand{\VAN}[3]{#2}
\let\VANthebibliography\thebibliography
\def\thebibliography{\DeclareRobustCommand{\VAN}[3]{##3}\VANthebibliography}

\usepackage{graphicx}	
\usepackage{amsmath}	

\title[Meet the Rotational Variables]{Seven Classes of Rotational Variables From a Study of 50,000 Spotted
Stars with ASAS-SN, Gaia, and APOGEE}

\author[A. Phillips et al.]{Anya Phillips$^{1}$\thanks{E-mail: phillips.1671@osu.edu},
C.~S. Kochanek$^{1,2}$,
Tharindu Jayasinghe$^{3,4}$,
Lyra Cao$^{1}$,
Collin T. Christy$^{5}$,
\newauthor D. M. Rowan$^{1}$,
and
Marc Pinsonneault$^{1}$
\\
$^{1}$Department of Astronomy, The Ohio State University, 140 West 18th Avenue, Columbus, OH, 43210, USA\\
$^{2}$Center for Cosmology and Astroparticle Physics, The Ohio State University, 191 W. Woodruff Avenue, Columbus, OH, 43210, USA\\
$^{3}$Department of Astronomy, University of California, Berkeley, CA 94720\\
$^{4}$NASA Hubble Fellow\\
$^{5}$Steward Observatory, University of Arizona, 933 North Cherry Avenue, Tucson, AZ 85721
}

\date{Accepted XXX. Received YYY; in original form ZZZ}

\pubyear{2023}

\begin{document}
\label{firstpage}
\pagerange{\pageref{firstpage}--\pageref{lastpage}}
\maketitle

\begin{abstract}
We examine the properties of $\sim50,000$ rotational variables from the ASAS-SN survey using distances, stellar properties, and probes of binarity from \textit{Gaia} DR3 and the SDSS APOGEE survey. They have high amplitudes and span a broader period range than previously studied \textit{Kepler} rotators. We find they divide into three groups of main sequence stars (MS1, MS2s, MS2b) and four of giants (G1/3, G2, G4s, and G4b). 
MS1 stars are slowly rotating (10-30 days), likely single stars with a limited range of temperatures.
MS2s stars are more rapidly rotating (days) single stars spanning the lower main sequence up to the Kraft break.
There is a clear period gap (or minimum) between MS1 and MS2s, similar to that seen for lower temperatures in the \textit{Kepler} samples.
MS2b stars are tidally locked binaries with periods of days.
G1/3 stars are heavily spotted, tidally locked RS CVn with periods of tens of days.
G2 stars are less luminous, heavily spotted, tidally locked sub-subgiants with periods of $\sim10$ days.
G4s stars have intermediate luminosities to G1/3 and G2, slow rotation periods (approaching 100 days) and are almost certainly all merger remnants.
G4b stars have similar rotation periods and luminosities to G4s, but consist of sub-synchronously rotating binaries.
We see no difference in indicators for the presence of very wide binary companions between any of these groups and control samples of photometric twin stars built for each group.
\end{abstract}




\section{Introduction}\label{section1}

Rotation provides a powerful stellar population diagnostic and is essential to understanding stellar structure and evolution. 
In stars with convective envelopes, rotationally-driven dynamos produce magnetic fields which in turn lead to starspots on the stellar surface (e.g., \citealt{Yadav2015}). If the star's rotation is fast enough and the spot fraction is large enough, the brightness of the star varies quasi-periodically, allowing a measurement of the rotation rate of these "rotational variables." Surface spot coverage is linked to mechanisms of interior angular momentum transport \citep{cao2023}, so studies of photometric modulation are well-suited to studying stellar structure and evolution.

Rapid rotation in low mass stars is traditionally regarded as an indicator of youth because the rotation rate in solar-mass stars, and their activity, decrease with age \citep{skumanich1972} due to angular momentum loss from magnetized solar-like winds \citep{weberdavis1967}. Lower mass stars take longer to spin down than Solar analogs, and are consistently more active at a given rotation period. Stellar activity is usually parameterized by a Rossby number, the ratio of the convective overturn timescale to the rotation period. High mass stars have much shorter overturn timescales than low mass stars, and are therefore inactive; this explains why low mass stars have magnetized winds and spin down, while higher mass stars do not \citep{durneylatour1978}. Stellar spin down is consequently a potentially important age indicator \citep{barnes2007}, especially in lower mass stars that experience little nuclear evolution in a Hubble time. 

We can model the correlation between rotation rate and age with gyrochronology, where the rotation rate of a  main sequence star is used as an age estimator. This method has blossomed with the large samples of low-amplitude rotational variables discovered by \textit{Kepler} \citep{keplerI2010, keplerII2010}. For example, \citet{McQuillan2014} derived rotation periods of $\sim34,000$ \textit{Kepler} main-sequence stars with amplitudes a low as 0.1\% and applied gyrochronological models to estimate their ages. In practice, it has proven challenging to quantify such gyrochronology relationships. For example, magnetic braking ceases in the oldest, least active stars \citep{vansaders2016}. There is also a transient phase where spin down pauses; this was first discovered in Solar analogs \citep{krishnamurthi1997}, but lasts for a longer time in K dwarfs \citep{curtis2019}, which complicates gyrochronology \citep{bouma2023}. 

Binary stars provide a completely different channel for inducing rapid rotation. Close binary systems are synchronized by tides, allowing low mass stars to remain active for their entire main sequence lifetime \citep{wilson1966}. Angular momentum lost in winds is extracted from the orbits of sufficiently short-period binaries, and this can produce mergers, sometimes referred to as blue stragglers, on the main sequence \citep{andronov2006}. 

Once off the main sequence, single stars expand and slow down, even without magnetized winds. As a result, most evolved giant stars are slow rotators. 
However, when mergers occur on the giant branch, the merger products can rotate rapidly. \citet{Daher2022} found that $0.8$--$3.5\%$ of $79,308$ APOGEE field giants in their sample rapidly rotate, depending on the chosen threshold for what constitutes "rapid rotation." Other studies (including \citealt{tayar2015}, \citealt{carlberg2011}, and \citealt{massarotti2008}) find rapid rotator fractions in this range with the exact values depending on varying amplitude thresholds and physical differences in the selected stellar populations \citep{Patton2023}. Many of these rapidly rotating giants are apparently single \citep{Patton2023} and are almost certainly merger products. 

Rapidly rotating giants can also result from tidal interaction in a binary, and giants in binary systems can become tidally synchronized at a wide range of periods (see \citealt{leiner2022} for a recent discussion). The combination of long overturn timescales and relatively short rotation periods (either due to tidal interaction or being a merger product) can produce extremely high activity in a minority of stars. 
Almost all magnetically active giants are therefore expected to either be merger products or currently interacting binary stars. 
\citet{Ceillier2017} found a high rate of interacting binaries and mergers on the red giant branch, showing 15\% of 575 low mass ($M<1.1M_{\odot}$) red clump stars from \textit{Kepler} to have detectable rotation through brightness modulations, inconsistent with single stars which are not merger products.
Further, \citet{gaulme2020} directly established the connection between rotational modulation due to starspots and tidal interaction for \textit{Kepler} red giants, finding $\sim85\%$ of non-oscillating red giants with rotational modulation to be in spectroscopic binaries.

Two known populations of rapidly rotating, synchronized binary giants are the RS Canum Venaticorum-type stars (RS CVn) \citep{hall1976}, and a less-luminous and shorter-period group of sub-subgiants \citep{leiner2022}. Both populations lie at the base of the giant branch. As giants become larger, the timescale for their evolution becomes shorter, while the timescale needed to synchronize increases \citep{verbuntphinney1995}. Fully synchronized systems are therefore not expected for luminous giants. However, merger products can appear at any luminosity.

In this paper, we carry out a population survey of rotational variables based on roughly 50,000 systems identified by the All-Sky Automated Survey for Supernovae (ASAS-SN) \citep{jayasinghe2018,jayasinghe2019I,jayasinghe2019II,jayasinghe2020, jayasinghe2021, christy2023}. These tend to be fairly high amplitude (10-30\%) and span a range of periods from $\sim$10 to 160 days (see Figure \ref{fig:kepler_comparison}). The key to our survey is the availability of distances through \textit{Gaia} \citep{gaia2016, gaia2021, gaia2022} and a broad range of stellar properties from both \textit{Gaia} and SDSS APOGEE DR17 \citep{apogeedr172022}. In particular, these supply considerable information on the binarity of systems. The starting point is the observation in \citet{christy2023}, for a version of the left panel of Figure \ref{fig:perlum+cmd} showing the distribution of ASAS-SN rotational variables in absolute magnitude and rotation period, that the rotational variables seemed to lie in discrete groups.
After dividing our sample, we examine each groups' detailed properties, in particular radial velocity variability, binarity, rotation rates, and spot coverage. We describe the data used in Section \ref{section2}, and then explore the properties of the empirically-divided groups in Section \ref{section3}. We conclude that there are seven distinct groups of rotational variables in Section \ref{section4} and discuss future directions.

\section{Observations and Methods}\label{section2}
\begin{figure*}
    \centering
    \includegraphics[width=\textwidth]{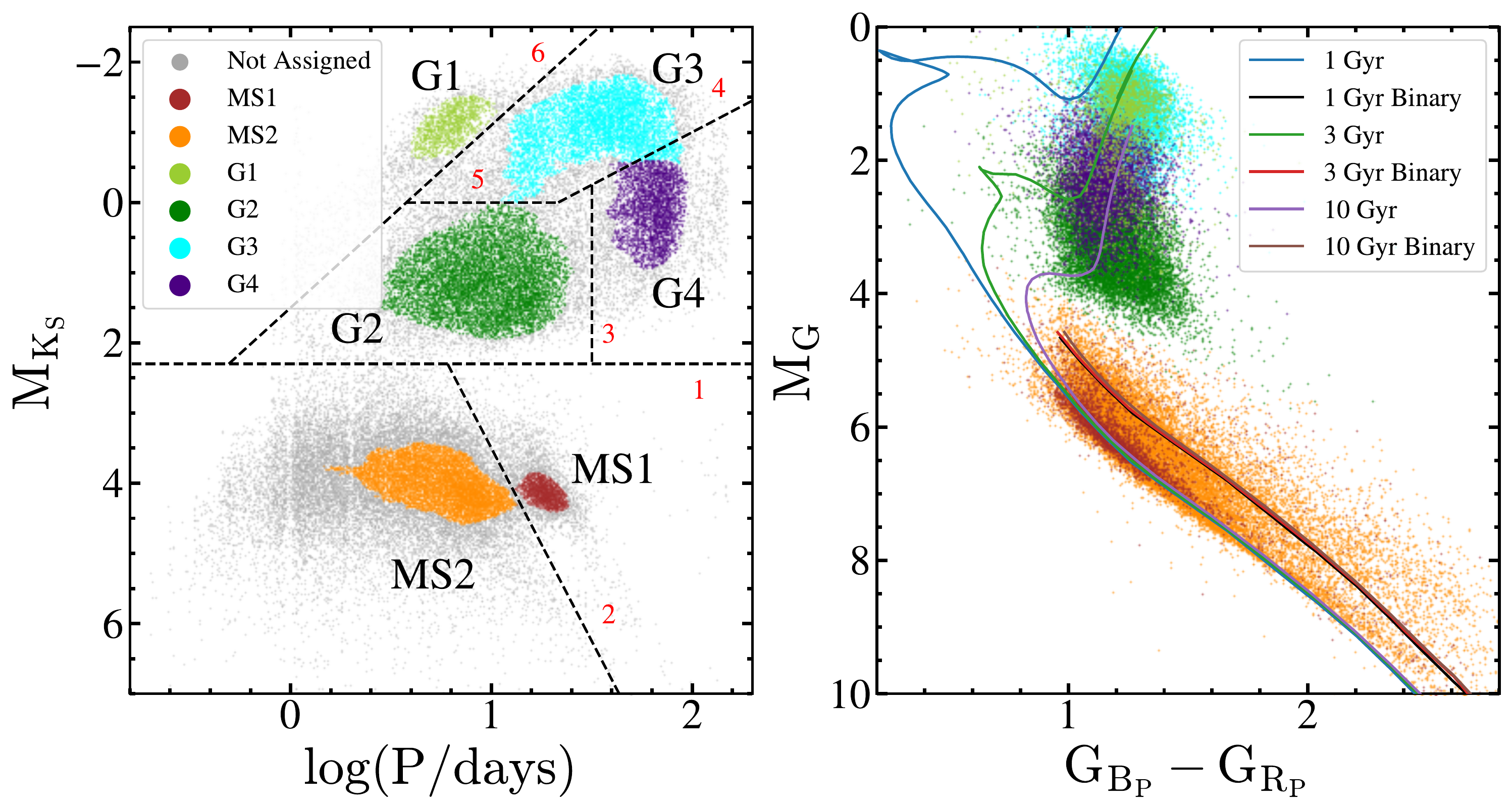}
    \caption{(Left panel) Distribution of the stars in absolute K$_s$ magnitude and period with the clusters assigned by \texttt{HDBSCAN}, and the boundaries used to manually assign all the stars to groups. The dashed boundaries are numbered in red, corresponding to those in Table \ref{tab:cluster_lines}. (Right panel) Distribution of the rotational variables in absolute G magnitude versus B$_P-$R$_P$ color, color-coded by the manually assigned cluster. We also include 1, 3, and 10 Gyr Solar metalicity PARSEC isochrones \citep{parsec1, parsec2}. The main sequence is also shown at twice the luminosity as the sequence of "twin" binaries. Stars above the sequence of twin binaries are probably young stellar objects (YSOs).}
    \label{fig:perlum+cmd}
\end{figure*}
\begin{figure*}
    \centering
    \includegraphics[width=\textwidth]{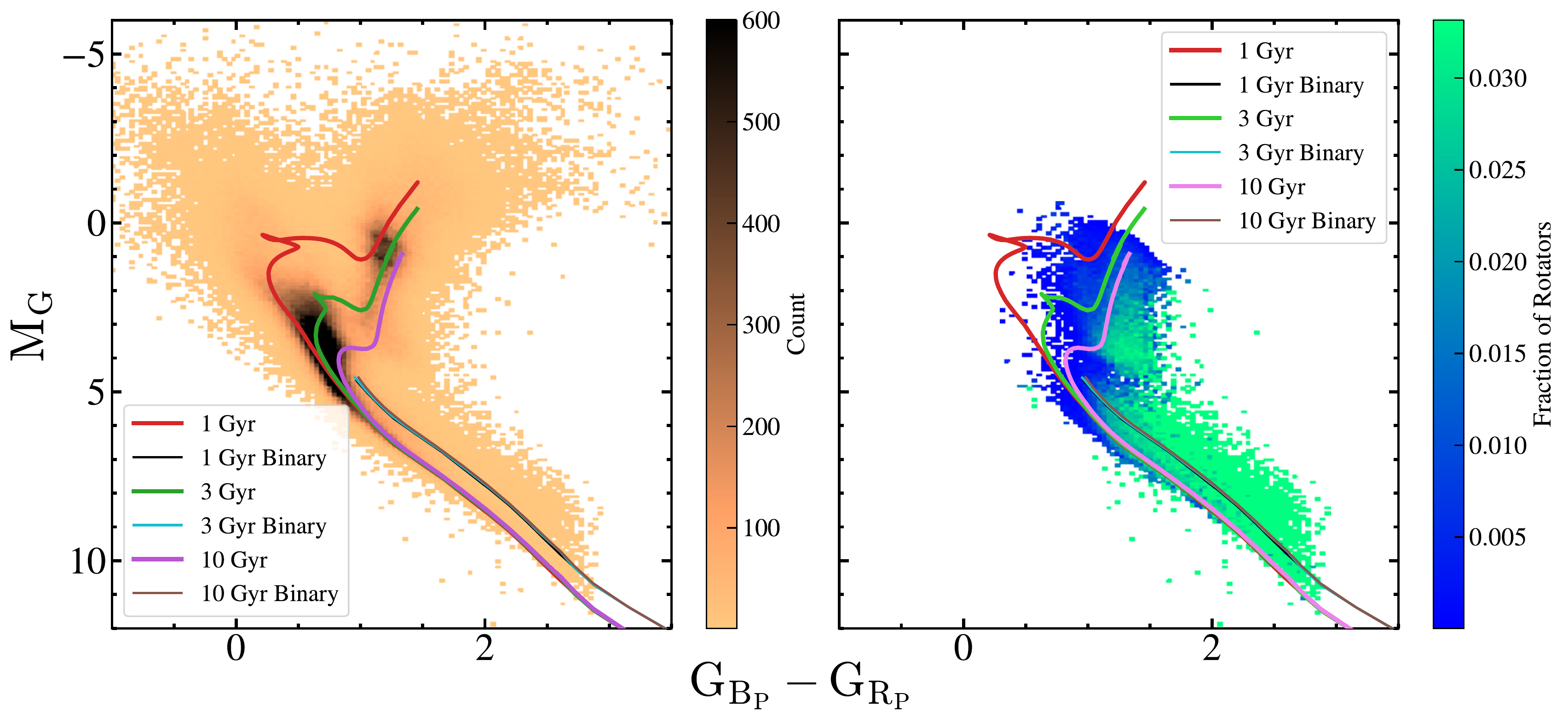}
    \label{fig:cmds}
    \caption{(Left panel) Absolute G magnitude versus $B_P-R_P$ color for a random sample of $\sim$500,000 ASAS-SN sources searched for variability where the bin color corresponds to the number of sources. (Right panel) Absolute G magnitude versus $B_P-R_P$ color where bin color corresponds to the fraction of stars which were identified as rotational variables. Both panels are overlayed with the same isochrones as in the right panel of Figure \ref{fig:perlum+cmd}.}
\end{figure*}
Here we consider the subsample of $48,298$ rotational variables shown in the left panel of Figure~\ref{fig:perlum+cmd}. We restricted the sample to systems with \textit{Gaia} EDR3 \citep{gaia2016, gaia2021} parallax signal-to-noise ratios of $>10$. We use distances from \citet{bailerjones2021} and extinction estimates from the \texttt{mwdust} \citep{bovy2016} `Combined19' dust map \citep{drimmel2003, marshall2005, green2019}. We keep systems with estimated extinctions $A_V<2$ and dispose of the small number of outliers with either extinction corrected $B_P-R_P>3$ or $M_G<-1$. The ASAS-SN variable catalog is dominated by the V band sample \citep{jayasinghe2018,jayasinghe2019I,jayasinghe2019II,jayasinghe2020, jayasinghe2021} which had significant systematic problems for periods near one day, so we reject systems with $\rm{log_{10}(Period/days)}$ between $-.0018$ and $.0005$. 

The left panel of Figure \ref{fig:perlum+cmd}, the distribution of our sample in period and absolute 2MASS \citep{2mass2006} K$_s$ magnitude, appears to have $6$ clusters; four of giants and two of main sequence stars. To test this more formally, we used the density-based clustering algorithm \texttt{HDBSCAN} \citep{hdbscan2017} to identify clusters in this parameter space. \texttt{HDBSCAN} assigns each source to one of the clusters or to noise. We first separated the main sequence and giants using the input parameter \texttt{min\_cluster\_size = 1000}. We further divided the main sequence using \texttt{min\_cluster\_size = 1000} with the additional parameters \texttt{min\_samples = 200}, and \texttt{cluster\_selection\_epsilon = .07}, and the giants using \texttt{min\_cluster\_size = 1000}, with the additional parameters \texttt{min\_samples = 150}, and \texttt{cluster\_selection\_epsilon = .07}. 

This combination of parameters lead to the identification of $7$ clusters, the $6$ identified by eye and a seventh associated with the 1 day period notch. We ignore this grouping (it is not shown in Figure~\ref{fig:perlum+cmd}) and assign it to the adjacent cluster. The combination of parameters we used in \texttt{HDBSCAN} were meant to maximize the number of points assigned to clusters, but nonetheless many of the stars were not assigned to any group, as can be seen in the left panel of Figure \ref{fig:perlum+cmd}. For our analysis, we divided the stars into 6 clusters using the lines shown in the left panel of Figure \ref{fig:perlum+cmd} and presented in Table \ref{tab:cluster_lines} to expand the \texttt{HDBSCAN} clusters to include all of the stars. 
We label these initial groups as MS1 and MS2 for the main sequence, G1 and G2 for the shorter period giants, and G3 and G4 for the longer period giants.
The right panel of Figure \ref{fig:perlum+cmd} shows these clusters in extinction-corrected absolute G magnitude and B$_P-$R$_P$ color and the groups also partially separate in this space. 
While we begin with these six groups identified in period and absolute magnitude, we find them to further subdivide using other parameters; we will discuss this in Section \ref{section3}.

We visually inspected 100 randomly selected light curves from each group. The light curves overwhelmingly are those of rotational variables with very little contamination. We had hoped that there would be some qualitative differences between the light curves of the different groups, but no such differences were apparent. The residual low-level contamination observed in the light curves and the fact that we do not expect our manual divisions to be perfect will lead to scatter in other parameter spaces. Nonetheless, these divisions suffice for our purpose of highlighting the bulk properties of each group.
We also checked distributions in ASAS-SN amplitudes for each group, but the only obvious trend is the selection effect that fainter stars need higher amplitudes to be identified as variables.

\begin{table*}
    \centering
    \caption{Boundaries used to separate groups (labeled in red in the left panel of Figure \ref{fig:perlum+cmd}).}
    \begin{tabular}{llr}
         Boundary \# & Equation                          & Range \\
         \hline
         1          & $\rm{M_{K_S} = 2.3}$                  & $\rm{-1 < \log(P/days) < 3}$  \\[1mm]
         2          & $\rm{M_{K_S} = 5.5\log(P/days)-2}$     & $\rm{0.782 < \log(P/days) < 3}$\\[1mm]
         3          & $\rm{\log(P/days) = 1.5}$          & $\rm{-0.25<M_{K_S}<2.3}$         \\[1mm]
         4          & $\rm{M_{K_S} = -1.5\log(P/days)+2}$    & $\rm{1.33<\log(P/days)<3}$    \\[1mm]
         5          & $\rm{M_{K_S} = 0}$                    & $\rm{-0.577<\log(P/days)<1.33}$\\[1mm]
         6          & $\rm{M_{K_S} = -2.6\log(P/days)+1.5}$  & $\rm{-0.308<\log(P/days)<3}$  \\
    \end{tabular}
    \label{tab:cluster_lines}
\end{table*}

We also extracted 500,000 random stars from the full sample of ASAS-SN stars searched for variability in \citet{christy2023}, and the left panel of Figure \ref{fig:cmds} shows their distribution in color and absolute magnitude. The right panel shows the fraction of sources identified as rotational variables. This uses the ratio of the number of rotational variables to the number of random sources statistically corrected to be the fraction of the full input sample.
This makes no attempt to determine selection effects, but there is a clear absence of rotational variables on the main sequence above the \citet{kraft1967} break and on the upper giant branch. Rotational variables are more common lower on the main sequence, along the binary main sequence and for the sub-subgiants \citep{leiner2022}.

\begin{figure}
    \centering
    \includegraphics[width=\columnwidth]{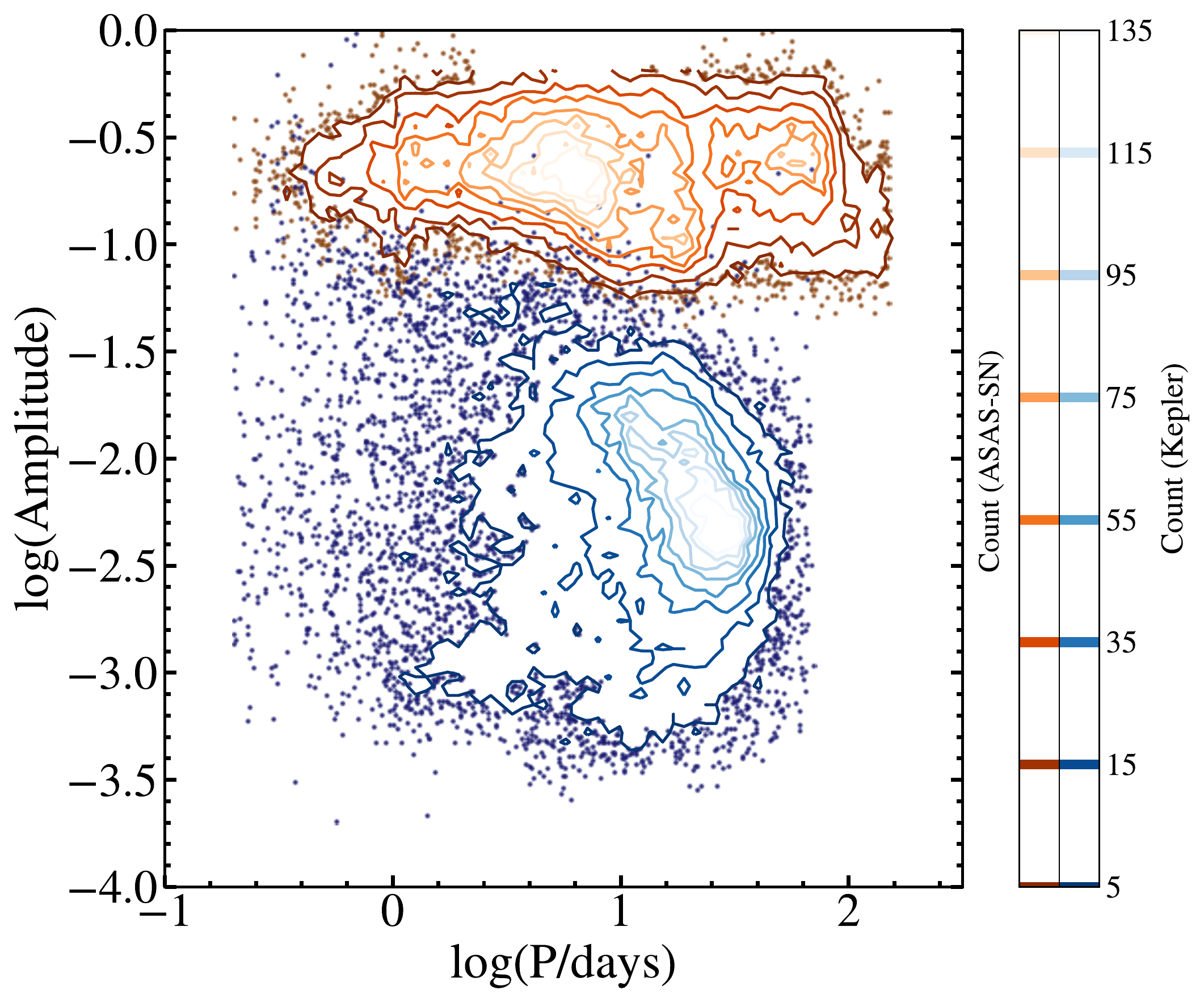}
    \caption{Distribution in amplitude and rotation period for the ASAS-SN (orange) and \textit{Kepler} (blue) rotational variables from \citet{McQuillan2014}. We binned the data in this parameter space and the lines show contours of constant bin count. 
    The amplitudes are in the g-band for ASAS-SN stars and in \textit{Kepler}'s bandpass for \textit{Kepler} stars. 
    \textit{Kepler} (ASAS-SN) amplitudes are the 5--95\% (2.5--97.5\%) ranges of the light curves.}
    \label{fig:kepler_comparison}
\end{figure}

Figure \ref{fig:kepler_comparison} compares our sample in amplitude and period to the \citet{McQuillan2014} sample of rotational variables in \textit{Kepler}. 
As a ground-based survey, ASAS-SN probes a higher-amplitude sample and over a broader period range than \textit{Kepler}, which focused on a field dominated by old stars with low variability amplitudes \citep{brown2011}. The two samples have essentially no overlap.

\begin{figure}
    \centering
    \includegraphics[width=\columnwidth]{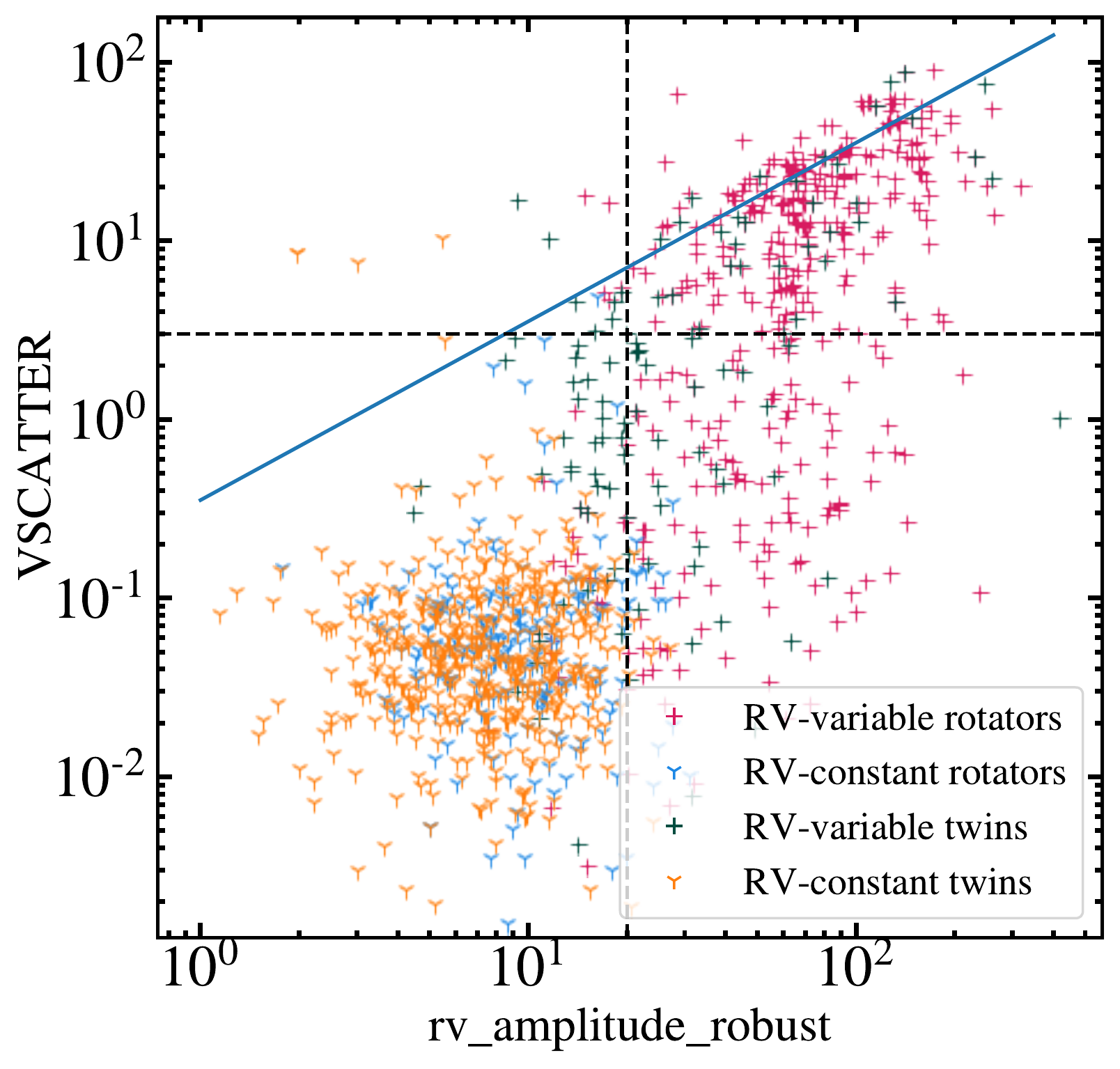}
    \caption{Comparison between APOGEE's \texttt{VSCATTER} parameter and \textit{Gaia}'s \texttt{rv\_amplitude\_robust}, including all rotators and twins meeting the criteria for the \textit{Gaia} RV-variability analysis and with APOGEE \texttt{NVISITS} $>1$, and coded by whether they meet the \citet{katz2022} criteria to be considered RV-variable. The blue diagonal indicates where \texttt{VSCATTER} $=\rm{\texttt{rv\_amplitude\_robust}}/2\sqrt{2}$. The vertical line indicates \texttt{rv\_amplitude\_robust}=20 km s$^{-1}$ and the horizontal line indicates \texttt{VSCATTER}=3 km s$^{-1}$.}
    \label{fig:vscatter_rvamp_comparison}
\end{figure}

We matched the rotational variables sample to APOGEE DR17 \citep{apogeedr172022} and \textit{Gaia} DR3 \citep{gaia2022}. Table \ref{tab:available_data} displays the number of stars in each survey as well as in certain subsets, and Table \ref{tab:frac_comparisons} shows the fractions of stars in each group with characteristics derived from this ancillary data. 
From \textit{Gaia},  we include in Table \ref{tab:frac_comparisons} the fraction of sources with RUWE $\geq$ 1.2, a conservative indicator for a wide binary or triple companion \citep{pearce2020, Belokurov2020}, flagged astrometric binaries (7- and 9-parameter acceleration solutions and astrometric orbits) \citep{astroorbit2022}, spectroscopic (SB1 and SB2) binaries \citep{SBI2022}, and systems with high dispersions in their \textit{Gaia} radial velocities (see below).  
The astrometric binaries, systems with astrometric accelerations, and systems with high RUWE are all associated with long period orbits that should not be directly associated with the rotational variability. We can see this explicitly in the astrometric binaries, where the typical period is 100-1000 days. Wide binaries can, however, indirectly be associated with the rotational variability if the system is really a triple and the long period companion drives the evolution of a short period inner binary
through Kozai-Lidov-type interactions (e.g., \citealt{Fabrycky2007}). 
We also use the Gaia $v_{\rm{broad}}$ parameter as an estimate of the stellar rotation $v \sin i$. Based on \citet{vbroad2022}, $v_{\rm{broad}}>10$ km s$^{-1}$ is indicative of fast rotation while lower values are consistent with noise. However, \citet{Patton2023} adopt the more conservative rapid rotation threshold  $v_{\rm{broad}}>20$ km s$^{-1}$, which yields better agreement with their rapid rotator fractions from APOGEE $v\sin i > 10$ km s$^{-1}$.

Gaia DR3 includes a number of variables which can be used to identify probable binaries through the scatter
of the individual radial velocity (RV) measurements compared to the estimated noise, as described in \citet{katz2022}. We
considered stars with
\begin{enumerate}
    \item \texttt{rv\_nb\_transits} $\geq$ 5,
    \item \texttt{rv\_expected\_sig\_to\_noise} $\geq$ 5, and
    \item 3900 $\leq$ \texttt{rv\_template\_teff} $\leq$ 8000.
\end{enumerate}
We found that either the \texttt{rv\_renormalised\_gof} or \texttt{rv\_amplitude\_robust} variables provided the clearest distinctions between probable binary and single (or wide binary) stars, where \texttt{rv\_renormalised\_gof} is a measure of the goodness of fit of a constant RV to the data, and \texttt{rv\_amplitude\_robust} is the peak-to-peak velocity amplitude after clipping outliers. \citet{katz2022} use a conservative criterion for a binary of $\texttt{rv\_renormalised\_gof}>4$ and $\texttt{rv\_chisq\_pvalue}\leq 0.01$.  For ease of comparison to the APOGEE \texttt{VSCATTER} (see below), we will focus on \texttt{rv\_amplitude\_robust}.  Based either on the value of \texttt{rv\_renormalised\_gof} or the comparison to APOGEE, we find that $\texttt{rv\_amplitude\_robust}\geq20$~km s$^{-1}$ is a good proxy for binarity. Requiring more transits (10) or higher signal-to-noise (10) had little effect on the results. 

We use the stellar parameters $T_{\rm{eff}}$, $\log g $ and $v \sin i$ from the APOGEE survey.  Where APOGEE has multiple RV measurements (\texttt{NVISITS}>1), we can use the root-mean-square 
scatter of the velocities \texttt{VSCATTER} as an indicator of binarity. 
Stars with \texttt{VSCATTER} $\geq 3$ km s$^{-1}$ (approximately equivalent to having a maximum difference in individual radial velocity measurements $\Delta RV_{\rm{max}}>$3--10 km s$^{-1}$) are almost certainly binaries \citep{badenes2018, Mazzola2020}. We also include the number of stars with estimates of starspot coverage from the LEOPARD spectroscopic analysis of \citet{cao2022}.  This algorithm fits the APOGEE spectrum using models with two different $T_{\rm{eff}}$ to estimate a temperature difference and a spot fraction $f_{\rm{spot}}$ for the fraction of the stellar surface associated with the cooler 
temperatures. This analysis can also interpret SB2's as spot contributions if the spectral types of each star are similar, which is only likely for similar mass main sequence binaries.

\begin{table*}
    \centering
    \caption{Statistics of available data.}
    \begin{tabular}{r|rrrrrrr}
        \hline
        \textbf{Rotators Sample}               & Total      & MS1  & MS2   & G1   & G2    & G3   & G4\\
        \hline
        All \textit{Gaia}                      &  48298     & 3258 & 20140 & 2342 & 10619 & 7201 & 4738\\ 
        Viable for \textit{Gaia} RV analysis                 &  38446     & 2860 & 14951 & 2177 & 7356  & 6867 & 4235\\
        All APOGEE                             &  2133      & 221  & 1073  & 64   & 395   & 219  & 161\\
        APOGEE \texttt{NVISITS} > 1            &  1438      & 139  & 711   & 44   & 273   & 156  & 115\\
        $f_{\rm{spot}}$ Estimates                   &  2121      & 219  & 1069  & 64   & 392   & 218  & 159\\
        
        \hline
        \textbf{Twins Sample}                  & Total      & MS1  & MS2   & G1   & G2    & G3   & G4\\
        \hline
        All \textit{Gaia}                      & 44836      & 3191 & 19665 & 2254 & 8224  & 6945 & 4557\\
        Viable for \textit{Gaia} RV analysis                 &   36879    &  2857& 14965 & 2100 & 6368  & 6559 & 4030\\
        All APOGEE                             & 1892       & 206  & 1039  & 66   & 188   & 238  & 155\\
        APOGEE \texttt{NVISITS} > 1            & 1279       & 132  & 694   & 44   & 114   & 178  & 117\\
        $f_{\rm{spot}}$ Estimates                   & 1873       & 203  & 1025  & 66   & 186   & 238  & 155\\
        \hline
    \end{tabular}
    \label{tab:available_data}
\end{table*}
\begin{table*}
    \centering
    \caption{Fractions of stars in each group with the characteristics listed in the left column. Fractions are for the subsample in Table \ref{tab:available_data} for which the measurement exists.}
    \label{tab:frac_comparisons} 
    \begin{tabular}{r|rrrrrr}

    \hline
    \textbf{Rotators Sample}                & MS1       & MS2      & G1      & G2      & G3       & G4\\
    \hline
        
    \textit{Gaia} RUWE $\geq$ 1.2 &     43.0\% &     35.0\% &     9.8\% &     12.2\% &     12.4\% &     12.6\% \\
    \textit{Gaia} Astrometric Binaries &     4.3\% &     3.5\% &     0.5\% &     0.5\% &     0.3\% &     0.7\% \\
    \textit{Gaia} Spectroscopic Binaries &     1.0\% &     0.8\% &     3.5\% &     1.4\% &     14.0\% &     9.9\% \\
    \textit{Gaia} Variable Radial Velocity &     13.9\% &     58.6\% &     98.4\% &     94.8\% &     87.1\% &     58.8\% \\
    \texttt{rv\_amplitude\_robust}$\geq 20$ km s$^{-1}$ &     13.3\% &     58.4\% &     97.1\% &     94.3\% &     79.7\% &     51.9\% \\
    APOGEE \texttt{VSCATTER} $\geq$ 3 km s$^{-1}$ &     4.3\% &     34.3\% &     86.4\% &     85.0\% &     73.7\% &     63.5\% \\

    \hline
    \textbf{Twin Sample}                    & MS1     & MS2     & G1      & G2      & G3      & G4\\
    \hline
    
    \textit{Gaia} RUWE $\geq$ 1.2 &     38.0\% &     43.1\% &     11.6\% &     15.4\% &     12.8\% &     14.6\% \\
    \textit{Gaia} Astrometric Binaries &     5.3\% &     6.3\% &     0.3\% &     1.2\% &     0.6\% &     1.2\% \\
    \textit{Gaia} Spectroscopic Binaries &     1.4\% &     0.7\% &     1.3\% &     0.8\% &     1.7\% &     1.5\% \\
    \textit{Gaia} Variable Radial Velocity &     14.7\% &     26.3\% &     11.5\% &     16.3\% &     10.6\% &     13.7\% \\
    \texttt{rv\_amplitude\_robust}$\geq 20$ km s$^{-1}$ &     13.7\% &     27.0\% &     6.4\% &     16.5\% &     5.3\% &     9.9\% \\
    APOGEE \texttt{VSCATTER} $\geq$ 3 km s$^{-1}$ &     10.6\% &     11.2\% &     4.6\% &     10.5\% &     1.7\% &     7.7\% \\
    
    \hline

    \end{tabular}
\end{table*}

To explore how the rotational variables compared to similar stars which are not known rotators, we constructed a sample of twins. For each star we selected all \textit{Gaia} stars with 
\begin{enumerate}
    \item 
        a parallax within 0.9 and 1.1 times the parallax of the rotator,
    \item 
        a difference in G magnitude less than $0.025n$,
    \item 
        a difference in B$_P$ magnitude less than $0.025n$, and
    \item 
        a difference in R$_P$ magnitude less than $0.025n$,
\end{enumerate}
where $n$ is an integer starting at $n=1$. We assign each star a metric which is simply the unweighted quadrature sum of the differences in parallax, $G$, $B_P$, and $R_P$ magnitudes and keep the lowest 16. If we find fewer than 16 stars, we iteratively increase $n$ until we have 16 stars. In most cases we succeed with $n=1$ and the overwhelming majority succeed for $n\leq2$. We then get the {\tt mwdust} extinction estimates for all 16 candidates and keep the one whose extinction is closest to the extinction of the rotator. We finally use the $\sim$93\% of twins whose extinctions agree to $|\Delta A_V|\leq0.2$ mag, which means that the extinction-corrected magnitudes and colors will have maximum differences due to the extinction mismatch of 0.2 and 0.1~mag, respectively.
By keeping more than 16 candidates we could still better match the extinctions, but this seemed good enough for our purposes given the small discrepancies in extinction-corrected photometry. We then extracted all of the ancillary data for the twins that we obtained for the rotators.  For all of the rotator classes except G2 this provided twins for $\simeq 97\%$ of the stars, while for G2 we are left with twins for only 77\% of the stars. Much of the G2 group lies brighter than the main sequence but redwards of the red giant branch. Such sub-subgiants are relatively rare, so it is not surprising that it is more difficult to find twins. 

The extinction-corrected absolute magnitude and color distributions of the twins and their corresponding rotational variables are very similar, as are their \textit{Gaia} $\log g$ and $T_{\rm{eff}}$ distributions. The APOGEE $\log g$ and $T_{\rm{eff}}$ distributions show several notable differences as can be seen from the summary statistics in Table \ref{tab:med_teff_logg}. The two MS samples are fairly similar, although the MS2 $T_{\rm{eff}}$ distribution of the twins extends to modestly (a few $100$~K) hotter temperatures.  
There are clear shifts for all of the giant groups, where the twins have systematically higher $T_{\rm{eff}}$ and lower $\log g$ than their corresponding rotators.  This is a known bias in the APOGEE parameters for active stars: APOGEE's analysis pipeline does not include rotation as a free parameter when fitting giant spectral templates, so the broadening of the lines created by spots and rotation in rapidly rotating giants strongly influence the derived stellar parameters, leading to underestimates of effective temperature and overestimates of surface gravity \citep{Patton2023}.
\begin{table}
    \centering
    \caption{Median APOGEE $\log g$ and $T_{\rm{eff}}$ for each group of rotators and its twin sample, with 16$^{\rm{th}}$ and 84$^{\rm{th}}$ percentile uncertainties.}
    \begin{tabular}{r|cc}
		      & $\log g$  & $T_{\rm{eff}}$ (K) \\
        \hline
        MS1 rotators & $4.57^{+0.03}_{-0.05}$ & $4735^{+233}_{-437}$\\[1mm]
        MS1 twins & $4.57^{+0.03}_{-0.04}$ & $4748^{+224}_{-364}$\\
        \hline
        MS2 rotators & $4.57^{+0.07}_{-0.16}$ & $4234^{+427}_{-390}$\\[1mm]
        MS2 twins & $4.59^{+0.05}_{-0.05}$ & $4431^{+375}_{-538}$\\
        \hline
        G1 rotators & $2.74^{+0.19}_{-0.13}$ & $4616^{+47}_{-77}$\\[1mm]
        G1 twins & $2.61^{+0.34}_{-0.19}$ & $4692^{+149}_{-106}$\\
        \hline
        G2 rotators & $3.57^{+0.36}_{-0.28}$ & $4435^{+197}_{-233}$\\[1mm]
        G2 twins & $3.42^{+0.27}_{-0.21}$ & $4901^{+156}_{-165}$\\
        \hline
        G3 rotators & $2.99^{+0.21}_{-0.30}$ & $4560^{+106}_{-121}$\\[1mm]
        G3 twins & $2.61^{+0.25}_{-0.14}$ & $4692^{+162}_{-120}$\\
        \hline
        G4 rotators & $3.30^{+0.23}_{-0.16}$ & $4682^{+321}_{-160}$\\[1mm]
        G4 twins & $3.12^{+0.26}_{-0.26}$ & $4821^{+162}_{-131}$\\
    \end{tabular}  
    \label{tab:med_teff_logg}
\end{table}

Figure \ref{fig:vscatter_rvamp_comparison} compares the APOGEE \texttt{VSCATTER} to the \textit{Gaia} \texttt{rv\_amplitude\_robust} for both the twins and the rotators. The points are coded by whether they meet the \citet{katz2022} criteria for RV variability.
For a sine wave, the radial velocity amplitude \texttt{rv\_amplitude\_robust}
would be $2\sqrt{2}$ larger than \texttt{VSCATTER}. The two estimates of velocity scatter are reasonably well correlated,  but the overall scatter is large because both are based on a small (APOGEE) or modest (\textit{Gaia}) number of measurements. Nonetheless, \texttt{rv\_amplitude\_robust}>20~km s$^{-1}$ is a reasonable proxy for binarity.

\section{Discussion}\label{section3}
\begin{figure*}
    \centering
        \includegraphics[width=.95\textwidth]{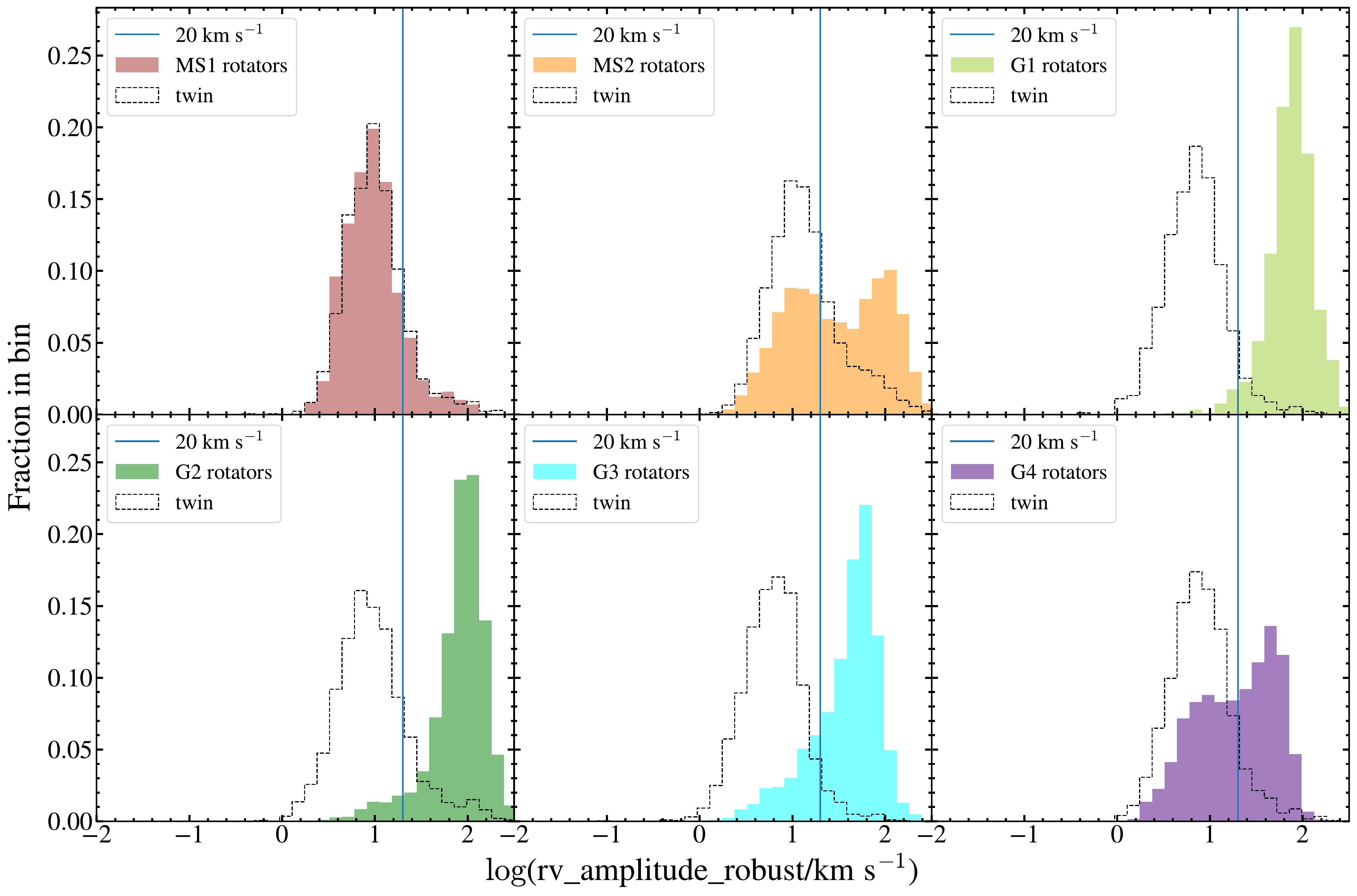}
        \includegraphics[width=0.95\textwidth]{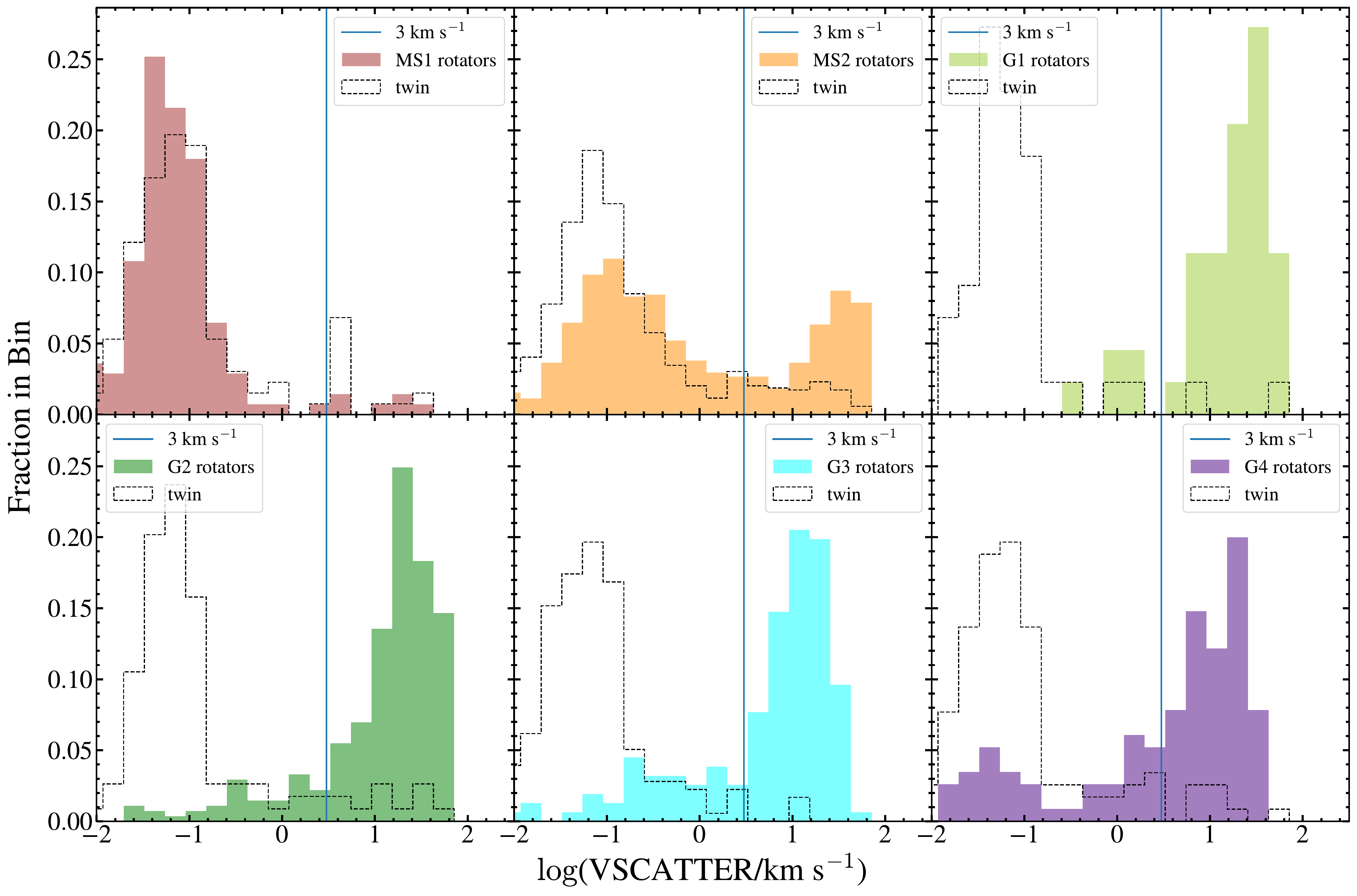}
    \caption{Distributions of each group of rotators and their twins in \textit{Gaia} \texttt{rv\_amplitude\_robust} (top) and in APOGEE \texttt{VSCATTER} (bottom). Stars to the right of the vertical lines at $20$ km/s (\textit{Gaia}) and $3$ km/s (APOGEE) are almost certainly binaries.}
    \label{fig:rvamp_twin_hists}
\end{figure*}
We expect binarity to play a key role in producing rotational variables, so we start with the distributions in \textit{Gaia} \texttt{rv\_amplitude\_robust} and APOGEE \texttt{VSCATTER}, shown for each group and its twin in Figure \ref{fig:rvamp_twin_hists}. We see that the MS1 distribution is single-peaked at low \texttt{rv\_amplitude\_robust} and \texttt{VSCATTER}, with a nearly identical distribution to its twins, and so MS1 consists largely of single stars (or sufficiently wide binaries). The rotators in MS2 are strongly bimodal with one group of single stars and one group of binaries which we label MS2s and MS2b, respectively. In the right panel of Figure \ref{fig:perlum+cmd}, we see that the MS2 group lies both on the main sequence (MS2s) and on the "binary main sequence," above the main sequence in magnitude where the luminosities of the PARSEC isochrones have been doubled (MS2b). The MS2 twins also show a significant tail of RV variables, almost certainly because twins of MS2b stars on the binary main sequence are also binaries.

Separating MS2s/b based only on radial velocity scatter yields a small sample, limited to stars with multiple radial velocity measurements, but we can also divide MS2 photometrically. 
We use the criteria from \citet{cao2022}, who fit a polynomial to the observed main sequence and defined photometric binaries as those at least 0.25 mag brighter than this fit. 
This method implies a binary fraction for MS2 of $\sim43\%$. For comparison, if we just split the APOGEE \texttt{VSCATTER} sample at 3 km s$^{-1}$, we would have a binary fraction of $\sim34$\%. Since this is incomplete because it does not account for binaries missed due to inclination, the two estimates are reasonably consistent.
We use the photometric division of MS2s/b in \S\ref{section4} when comparing our main sequence sample to that of \citet{McQuillan2014}.

\begin{figure}
    \centering
    \includegraphics[width=\columnwidth]{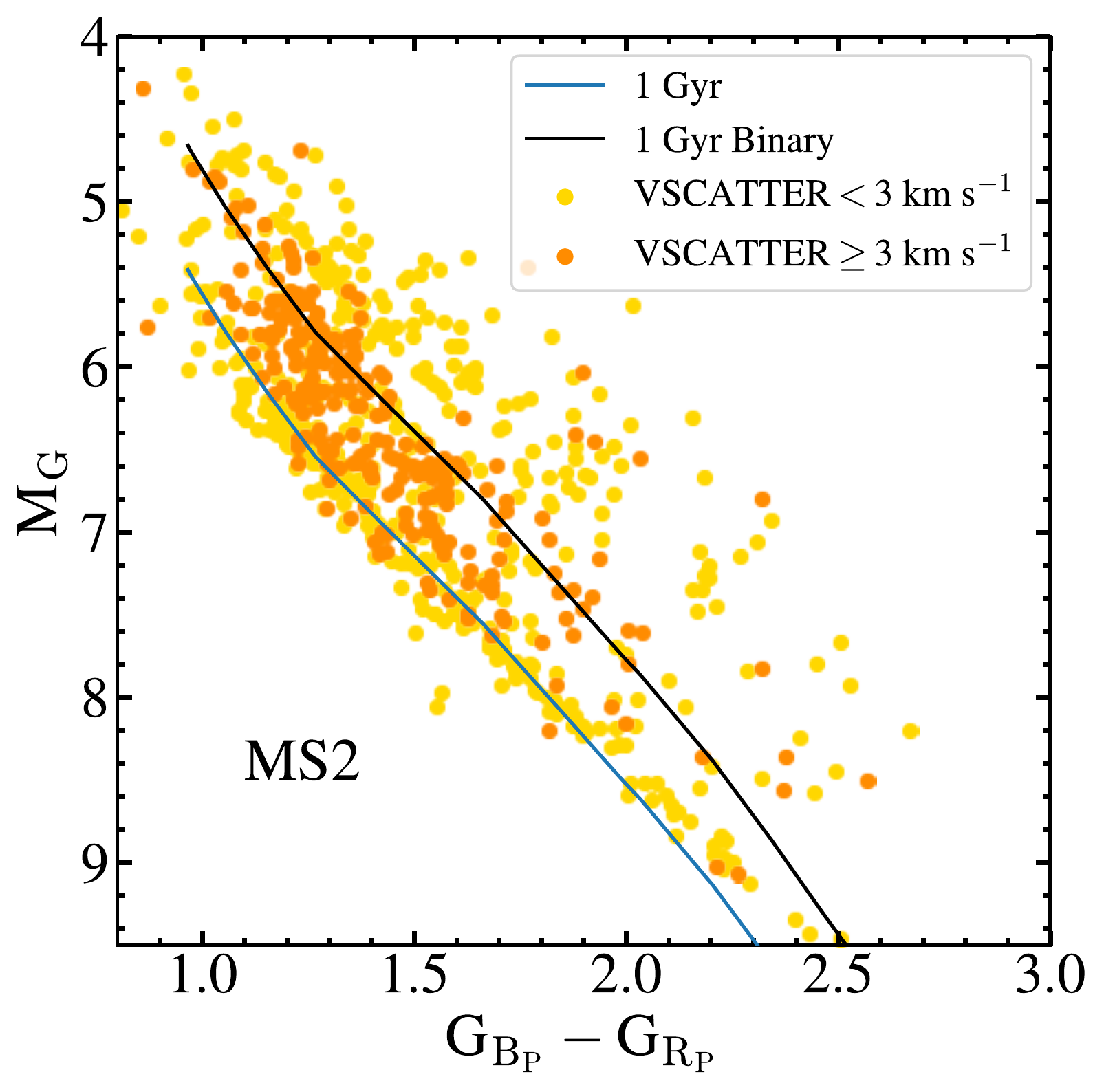}
    \caption{Absolute G magnitude versus $\rm{B_P-R_P}$ color for MS2 stars with valid \texttt{VSCATTER} measurements with the single and binary 1 Gyr PARSEC isochrones from Figure \ref{fig:perlum+cmd} overlayed.}
    \label{fig:ms2_cmd}
\end{figure}

To compare the results of separating MS2s/b photometrically and based on radial velocity scatter, Figure \ref{fig:ms2_cmd} shows a color-magnitude diagram of MS2 stars colored by whether \texttt{VSCATTER}$\geq3$ km s$^{-1}$. We see a band of low-\texttt{VSCATTER} stars on the main sequence, a band of high-\texttt{VSCATTER} stars on the binary main sequence, and a third band of low-\texttt{VSCATTER} stars above the binary main sequence. This upper band of stars with low RV scatter is not present in color-magnitude diagrams where we split instead by whether \texttt{rv\_amplitude\_robust}$\geq20$ km s$^{-1}$, or based on RV variability according to criteria from \citet{katz2022}.
Unresolved triple systems could lie above the binary main sequence without having significant radial velocity scatter, but we saw no indication in \textit{Gaia} RUWE that this upper band of low-\texttt{VSCATTER} stars are triple systems. Instead, these are probably young stellar objects (YSOs), which also have quasi-periodic rotational modulation, but lie well above the main sequence \citep{rebull2016}. To confirm this, we verified that that the YSO candidates have higher than average $A_V$.

The G1-G3 rotators are all clearly binaries based on their \texttt{rv\_amplitude\_robust} and \texttt{VSCATTER} distributions, whereas their twins are predominantly single stars. While the G4 twins are overwhelmingly single, the G4 distribution is bimodal, indicating subpopulations of both single (G4s) and binary (G4b) stars. Note that the bimodality seen in MS2 and G4 is real and not due to inclination. Inclination effects produce distributions with the rapidly dropping tails to lower velocity seen for G3.
Formally, for a true binary of orbital velocity $v_T$, the observed velocity $v_0$ is distributed as $x(1-x^2)^{-1/2}$, with $x=v_0/v_T$ for a uniform distribution in $\cos i$.
The single star subset of G4 likely consists of merger products. The distributions in \textit{Gaia} \texttt{rv\_renormalised\_gof} confirm the results of the \texttt{rv\_amplitude\_robust} and \texttt{VSCATTER} distributions, in particular the existence of the G4 merger subpopulation. 

\begin{figure*}
    \centering
        \includegraphics[width=0.95\textwidth]{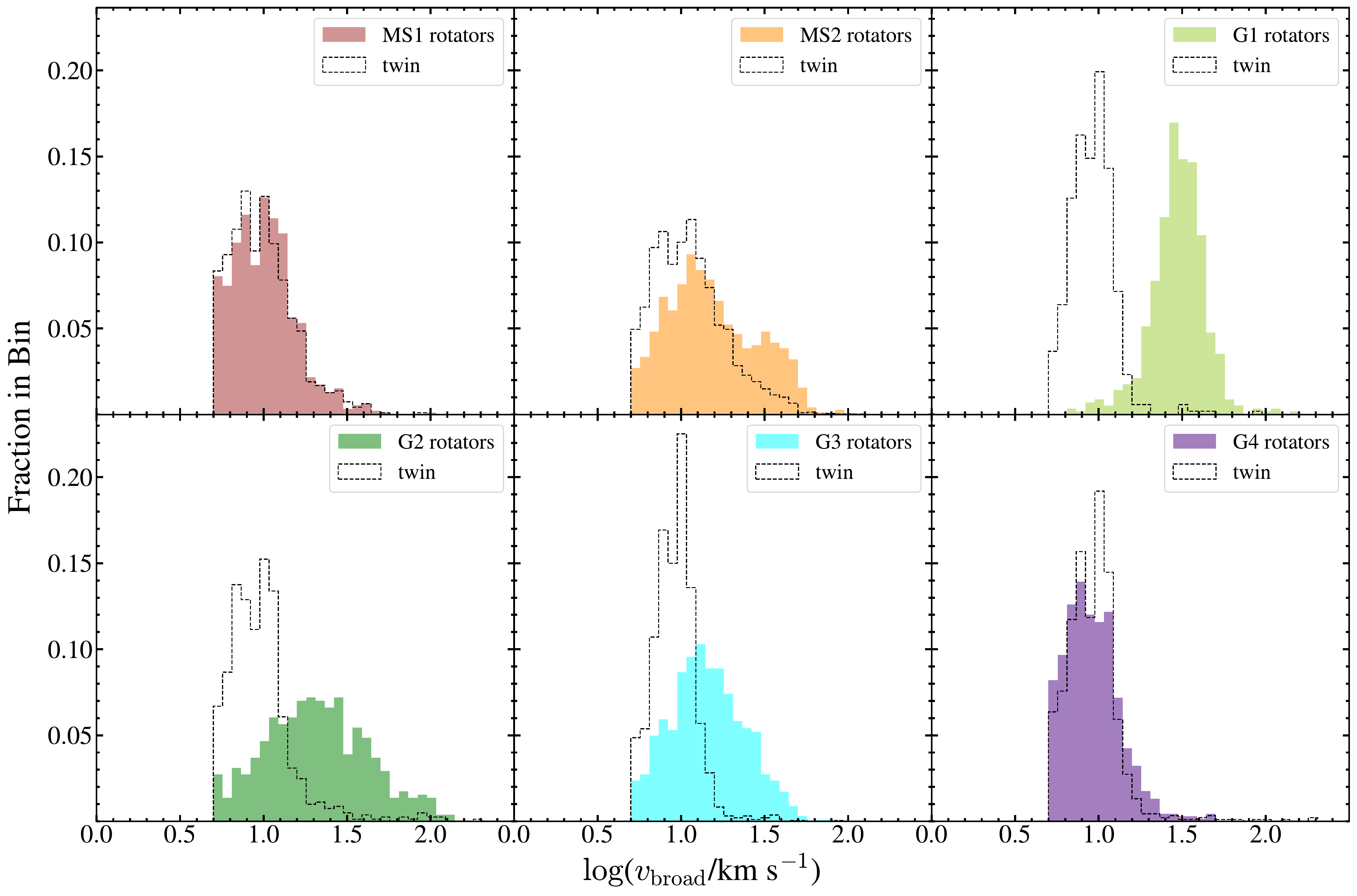}
        \includegraphics[width=0.95\textwidth]{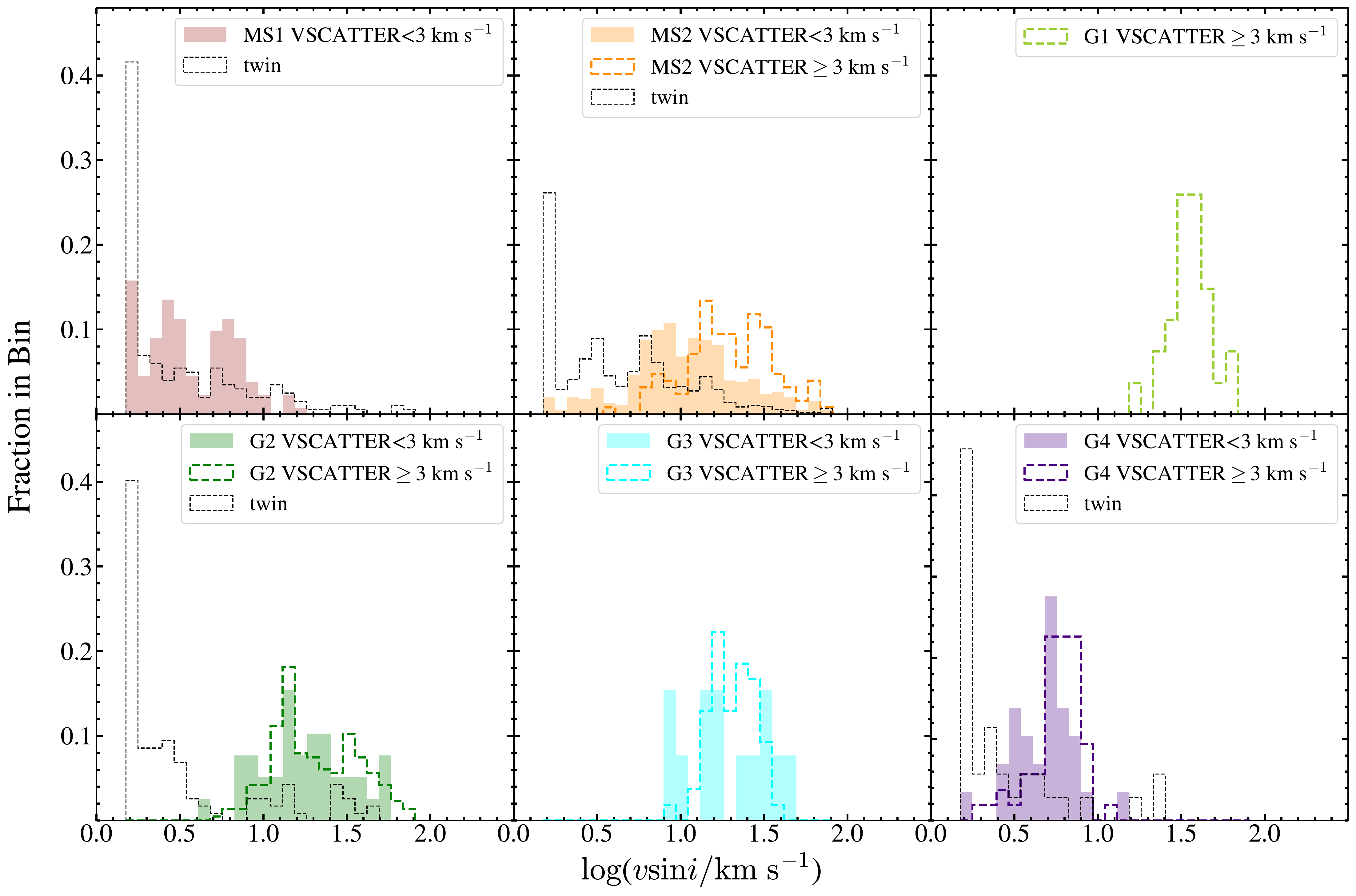}

    \caption{Distributions in $v_{\rm{broad}}$ (top) and  $v\sin i$ (bottom) for each group and their twins. The $v\sin i$ distributions are split by \texttt{VSCATTER}, where the solid histograms are for \texttt{VSCATTER}<3 and the dashed histograms are for \texttt{VSCATTER}$\geq$3 km s$^{-1}$. The G1 and G3 twins, as well as the high-\texttt{VSCATTER} population of MS1 and low-\texttt{VSCATTER} population of G1 have $<10$ APOGEE $v\sin i$ measurements and are not shown because there are too few systems.}
    \label{fig:vsini_vbroad_twin_hists}
\end{figure*}
\begin{figure*}
    \centering
    \includegraphics[width=\textwidth]{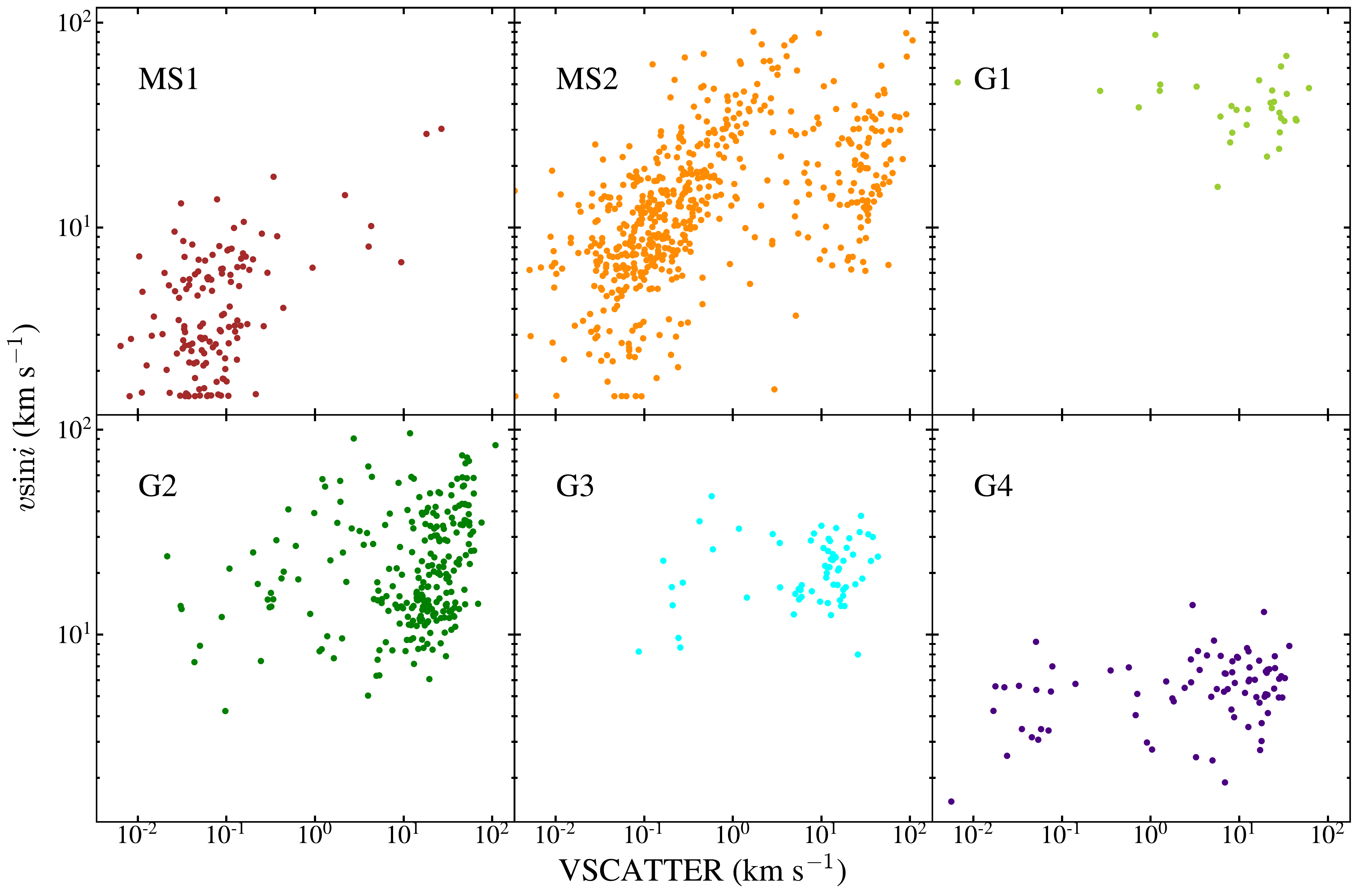}
    \caption{Distributions in APOGEE $v\sin i$ and \texttt{VSCATTER} for each group.}
    \label{fig:vsini_vscatter}
\end{figure*}
We also expect rotation rates to be an important physical probe of rotational variables, and Figure \ref{fig:vsini_vbroad_twin_hists} shows the twin and rotator distributions in APOGEE $v\sin i$ and \textit{Gaia} $v_{\rm{broad}}$,  as an estimator of $v\sin i$. The $v\sin i$ distributions include only stars with \texttt{NVISITS}>1 and are split by whether \texttt{VSCATTER}$\geq 3$ km s$^{-1}$. The median $v\sin i$ values for each subgroup are given in Table \ref{tab:med_fspot}. While the APOGEE  $v\sin i$ should be accurate even for small velocities, comparisons with APOGEE show that values of \textit{Gaia} $v_{\rm{broad}}<10$~km s$^{-1}$ should be regarded as upper limits \citep{vbroad2022}.
MS1, and its twin are comprised of slow rotators. In Table \ref{tab:med_fspot}, we see that the high \texttt{VSCATTER} stars of MS1 have a much higher median $v\sin i$, but the small number of them means that they are not included in Figure \ref{fig:vsini_vbroad_twin_hists}.
MS2, particularly in $v_{\rm{broad}}$, again seems bimodal, while its twin is comprised predominantly of slower rotators.
In Figure~\ref{fig:vsini_vscatter}, where we show the distribution of MS1 and MS2 in both \texttt{VSCATTER} and $v\sin i$, there is a clear separation of MS2 into two populations, where the high-\texttt{VSCATTER} systems all have high $v\sin i$. MS2's bimodality is less obvious in the $v\sin i$ distributions because there are low-\texttt{VSCATTER} systems with high $v\sin i$. The slower MS2 rotators still seem to have larger $v\sin i$ than the MS2 twins and MS1. 

\begin{figure*}
    \centering
    \includegraphics[width=\textwidth]{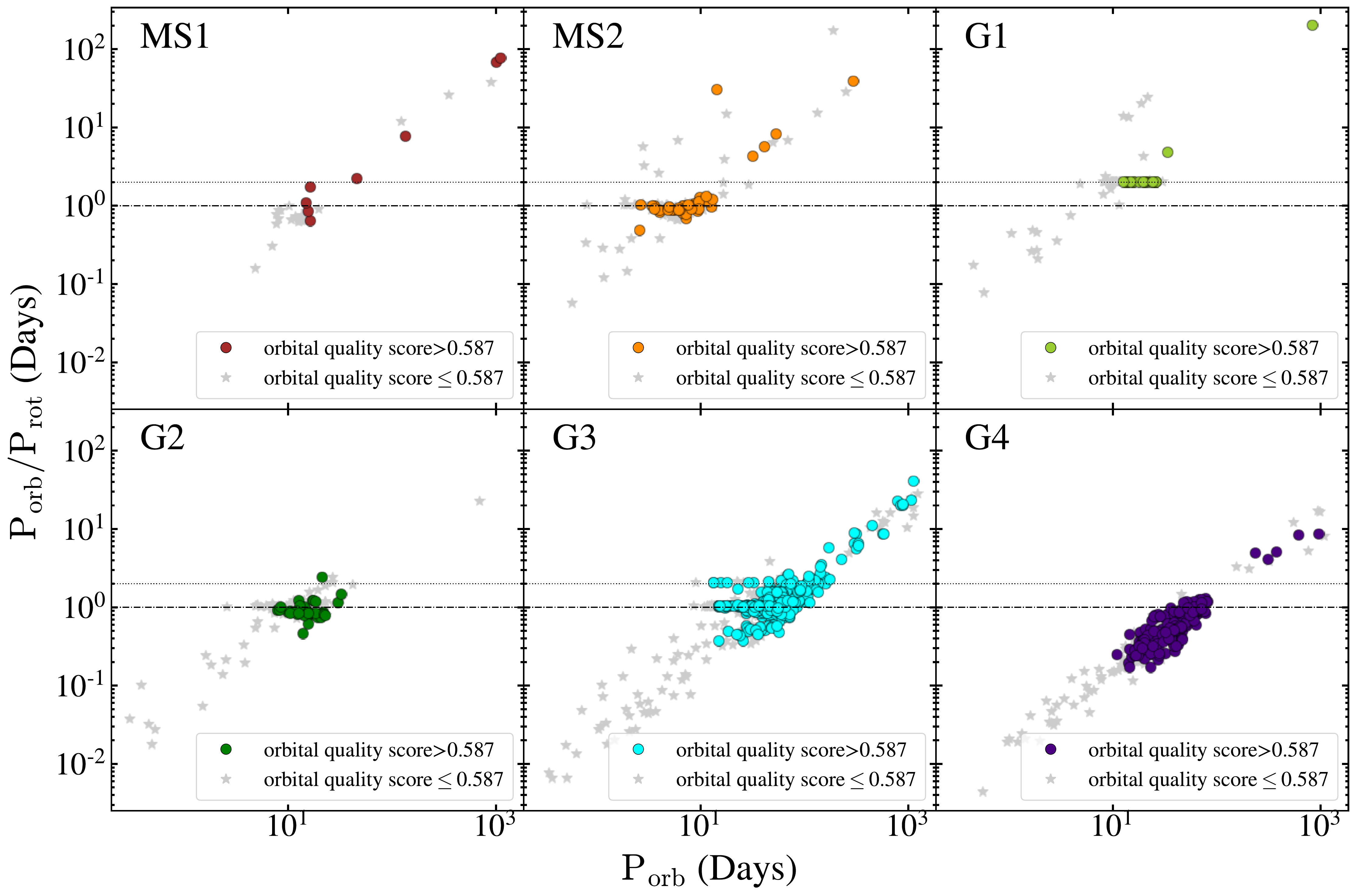}
    \caption{Ratio of the \textit{Gaia} SB1 $P_{\rm{orb}}$ to the ASAS-SN $P_{\rm{rot}}$ versus $P_{\rm{orb}}$. The lower dashed line is for $P_{\rm{orb}}=P_{\rm{rot}}$ and the upper dotted line is for $P_{\rm{orb}}=2P_{\rm{rot}}$. The colorful points have orbital scores $>0.587$ and the grey background points have orbital scores $\leq0.587$.}
    \label{fig:sbbinary_rotation_period}
\end{figure*}

The giants are in three groups. G1 has high $v\sin i$ and $v_{\rm{broad}}$, and while its twin group has very few $v\sin i$ measurements, it tends to have lower $v_{\rm{broad}}$. 
G4 has the slowest rotation rates of the giants, and while it has a similar $v_{\rm{broad}}$ distribution to its twin, the G4 twins have still slower $v\sin i$. Of the giant groups, G4 is the only one where the high-\texttt{VSCATTER} stars do not have a high $v\sin i$.  The G4 stars have long rotational periods, so the smaller $v \sin i $ are expected, but the cause is physically
interesting.  The G4 binary stars are not tidally locked, but are in sub-synchronous orbits (see the discussion associated with Figure \ref{fig:sbbinary_rotation_period}).
G2 and G3 have intermediate rotation rates to G1 and G4, and their twins tend to have slower rotation rates (though the G3 twins have very few $v\sin i$ measurements).
Note that because of the crudeness of the $v_{\rm{broad}}$ parameter, the \textit{Gaia} equivalent of Figure \ref{fig:vsini_vscatter}, $v_{\rm{broad}}$ versus \texttt{rv\_amplitude\_robust}, is uninformative.

\begin{figure*}
   \centering
    \includegraphics[width=\textwidth]{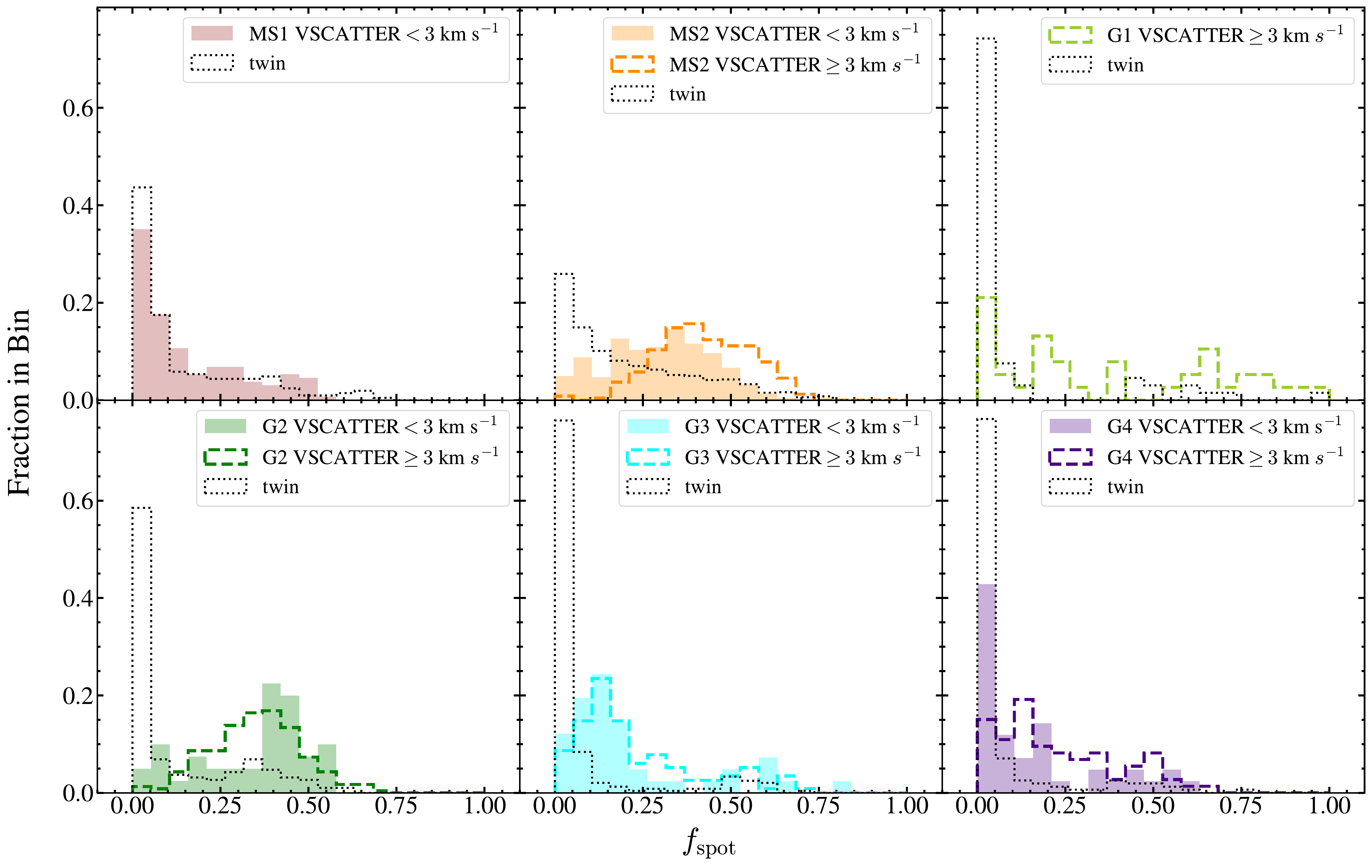}
    \caption{Distributions of the spot fraction $f_{\rm{spot}}$ for each group. The solid histograms are for rotational variables with APOGEE \texttt{VSCATTER} < 3 km s$^{-1}$ and the colorful dashed histograms are distributions for stars with \texttt{VSCATTER}$\geq$3 km s$^{-1}$. The black dashed histograms are for the twin groups. Only distributions with $\geq10$ members are shown.}
    \label{fig:fspot_scattersplit}
\end{figure*}
Figure \ref{fig:fspot_scattersplit} shows the distributions of the stars in the  starspot filling fraction ($f_{\rm{spot}}$) where the rotational variables are also divided into likely binaries with $\texttt{VSCATTER} \geq 3$ and likely non-binaries with $\texttt{VSCATTER} < 3$.  Cases when there are fewer than $10$ stars are not shown.   Table~\ref{tab:med_fspot} gives the median spot fractions of each group.  Except for the MS2 twins, the twin populations have $f_{\rm{spot}}$ distributions strongly peaked near zero.  It seems likely that many of the tails of the twin distributions towards higher spot fractions are due to the presence of spotted stars which have not been recognized as rotational variables.

The spot fractions are estimated from the individual APOGEE epochs, and so are unaffected by the presence of orbital Doppler shifts, but they can be affected by the presence of spectral contamination from the companion. This means that only the main sequence binary sub-population MS2b, which tends to lie close to the binary main sequence, can have significant biases in $f_{\rm{spot}}$ due to the presence of a binary companion.  The giant binaries will generally have a much lower luminosity main sequence companion which cannot significantly contaminate the giant's spectrum.  

MS1 contains few binaries and has a spot fraction distribution nearly identical to its twins.   There is, however, a modest binary sub-population that is more heavily spotted (see Table \ref{tab:med_fspot}).
MS2 has significantly higher spot fractions than its twins, and the non-binary MS2s sub-sample is modestly less spotted than the binary MS2b sub-sample.  The MS2s stars are, however, much more spotted than the MS1 stars supporting the argument than they are two different populations rather than a continuum. 
Here the long tail on the twin distribution is  due to the tendency of MS2b twins to also be binaries. We confirmed that the high-$f_{\rm{spot}}$ MS2 twins also tend to have high \texttt{VSCATTER}.  

G1 has a very broad range of spot fractions, although the median of $f_{\rm{spot}} \sim 28\%$ is less than that of the MS2 and G2 groups and similar to that of the MS1 group.   Except for the likely merger sub-population of G4, the spotted giants are all binaries based on their  \texttt{VSCATTER} distributions.  This means that the low-\texttt{VSCATTER} systems should be dominated by binaries viewed at high inclinations rather than being a physically distinct population, so we expect similar spot fractions at high- and low-\texttt{VSCATTER} for G1, G2 and G3 but not G4, as we see in Figure \ref{fig:fspot_scattersplit} and Table~\ref{tab:med_fspot}.
The G2 group has the highest median spot fractions with the low-\texttt{VSCATTER} group having a modestly higher median ($f_{\rm{spot}} \simeq 40\%$ versus $36\%$). 
The G3 group has some of the lowest spot fractions with little difference between the high- and low-\texttt{VSCATTER} sub-samples. Like G2, the \texttt{VSCATTER} distributions again argue for a purely binary population.
Finally, the G4 group has some of the lowest spot fractions, but the high-\texttt{VSCATTER} systems are significantly more spotted than the low-\texttt{VSCATTER} systems. Unlike G1, G2, and G3, G4 does have a large number of non-binary, low-\texttt{VSCATTER} systems, so it is not surprising that they also have different spot fractions. 
\begin{table*}
    \centering
    \caption{Median $f_{\rm{spot}}$, $v\sin i$, $v_{\rm{broad}}$, $P_{\rm{rot}}$, and $P_{\rm{orb}}$ of each group and their twin as a whole, and each group of rotators separated by whether \texttt{VSCATTER}$\geq3$ km s$^{-1}$ (for stars with >1 APOGEE visit) or whether \texttt{rv\_amplitude\_robust}$\geq20$ km s$^{-1}$ (for stars meeting the criteria for \textit{Gaia} radial velocity analysis outlined in Section \ref{section2}). For some subsets, there were no stars with orbital period estimates from \textit{Gaia}, so their $P_{\rm{orb}}$ entries are left blank. Stars from the "twin" sample are not ASAS-SN rotational variables and do not have ASAS-SN rotational periods, so their $P_{\rm{rot}}$ entries are marked "n/a."}
    \begin{tabular}{r|rrrrr}
    median:                                    & $f_{\rm{spot}}$ & $v\sin i$ (km/s) & $v_{\rm{broad}}$ (km/s) & $P_{\rm{rot}}$ (days)  & $P_{\rm{orb}}$ (days)\\
    \hline
    All MS1 Rotators &      0.09 &       3.4 &       9.9 &      18.1 &      13.9 \\
    MS1 \texttt{VSCATTER}$<3$ km s$^{-1}$ &      0.09 &       3.3 &       9.4 &      18.1 &        \\
    MS1 \texttt{VSCATTER}$\geq3$ km s$^{-1}$ &      0.28 &      10.2 &      24.5 &      12.7 &        \\
    MS1 \texttt{rv\_amplitude\_robust}$<20$ km s$^{-1}$ &      0.06 &       2.7 &       9.8 &      18.5 &     180.6 \\
    MS1 \texttt{rv\_amplitude\_robust}$\geq20$ km s$^{-1}$ &      0.27 &       7.7 &      12.2 &      16.7 &      14.1 \\
    MS1 twins &      0.06 &       2.2 &       9.3 &       n/a &      25.3 \\
    \hline
    All MS2 Rotators &      0.35 &      12.8 &      14.1 &       3.7 &       6.1 \\
    MS2 \texttt{VSCATTER}$<3$ km s$^{-1}$ &      0.30 &      10.6 &      15.0 &       5.0 &        \\
    MS2 \texttt{VSCATTER}$\geq3$ km s$^{-1}$ &      0.41 &      20.2 &      15.0 &       2.3 &       6.1 \\
    MS2 \texttt{rv\_amplitude\_robust}$<20$ km s$^{-1}$ &      0.17 &       6.6 &      11.9 &       8.5 &       4.0 \\
    MS2 \texttt{rv\_amplitude\_robust}$\geq20$ km s$^{-1}$ &      0.37 &      16.5 &      20.2 &       3.3 &       6.1 \\
    MS2 twins &      0.14 &       3.6 &      10.7 &       n/a &      23.8 \\
    \hline
    All G1 Rotators &      0.24 &      37.8 &      30.2 &       5.4 &      13.4 \\
    G1 \texttt{VSCATTER}$<3$ km s$^{-1}$ &      0.27 &      48.1 &      41.5 &       7.9 &        \\
    G1 \texttt{VSCATTER}$\geq3$ km s$^{-1}$ &      0.28 &      36.3 &      32.3 &       5.2 &       9.6 \\
    G1 \texttt{rv\_amplitude\_robust}$<20$ km s$^{-1}$ &      0.31 &      48.7 &      34.3 &      10.2 &        \\
    G1 \texttt{rv\_amplitude\_robust}$\geq20$ km s$^{-1}$ &      0.16 &      37.8 &      30.0 &       5.8 &      13.4 \\
    G1 twins &      0.01 &       2.2 &       9.2 &       n/a &     240.6 \\
    \hline
    All G2 Rotators &      0.37 &      20.0 &      20.3 &       9.0 &      12.4 \\
    G2 \texttt{VSCATTER}$<3$ km s$^{-1}$ &      0.40 &      18.1 &      13.2 &       8.1 &        \\
    G2 \texttt{VSCATTER}$\geq3$ km s$^{-1}$ &      0.36 &      19.2 &      27.1 &       9.0 &      13.1 \\
    G2 \texttt{rv\_amplitude\_robust}$<20$ km s$^{-1}$ &      0.15 &       8.5 &      11.6 &      23.5 &     695.2 \\
    G2 \texttt{rv\_amplitude\_robust}$\geq20$ km s$^{-1}$ &      0.35 &      19.7 &      21.1 &       9.0 &      12.3 \\
    G2 twins &      0.02 &       2.1 &       9.1 &       n/a &      35.7 \\
    \hline
    All G3 Rotators &      0.15 &      21.3 &      14.0 &      30.9 &      41.1 \\
    G3 \texttt{VSCATTER}$<3$ km s$^{-1}$ &      0.14 &      17.9 &      13.3 &      41.9 &      48.9 \\
    G3 \texttt{VSCATTER}$\geq3$ km s$^{-1}$ &      0.16 &      21.2 &      17.0 &      29.4 &      36.0 \\
    G3 \texttt{rv\_amplitude\_robust}$<20$ km s$^{-1}$ &      0.13 &      15.2 &      11.7 &      49.3 &      45.4 \\
    G3 \texttt{rv\_amplitude\_robust}$\geq20$ km s$^{-1}$ &      0.14 &      23.2 &      14.8 &      32.0 &      41.1 \\
    G3 twins &      0.01 &      20.0 &       9.3 &       n/a &     352.7 \\
    \hline
    All G4 Rotators &      0.13 &       5.4 &       9.0 &      64.4 &      28.9 \\
    G4 \texttt{VSCATTER}$<3$ km s$^{-1}$ &      0.10 &       5.3 &       9.5 &      64.9 &      26.2 \\
    G4 \texttt{VSCATTER}$\geq3$ km s$^{-1}$ &      0.17 &       6.0 &       7.2 &      59.8 &      36.9 \\
    G4 \texttt{rv\_amplitude\_robust}$<20$ km s$^{-1}$ &      0.01 &       3.5 &       8.9 &      74.8 &      17.4 \\
    G4 \texttt{rv\_amplitude\_robust}$\geq20$ km s$^{-1}$ &      0.14 &       5.8 &       9.0 &      66.8 &      29.7 \\
    G4 twins &      0.00 &       1.9 &       9.1 &       n/a &     175.5 \\
    \hline
    \end{tabular}
    \label{tab:med_fspot}
\end{table*}

Figure \ref{fig:sbbinary_rotation_period} shows the ratio of the \textit{Gaia} SB1 orbital period, $P_{\rm{orb}}$, to the ASAS-SN rotational period, $P_{\rm{rot}}$, as a function of $P_{\rm{orb}}$ for each group. We expect many rotational variables to be tidally synchronized binaries so we include lines at $P_{\rm{orb}}=P_{\rm{rot}}$ and $P_{\rm{orb}}=2P_{\rm{rot}}$. There is no physics that would yield the $P_{\rm{orb}}=2P_{\rm{rot}}$ ratio, but it is difficult to measure an orbital period to be half it's actual value. Therefore, rotators lying on this line likely have a reported rotational period aliased to half the true value (see below). 

Many of the widely-scattered points are likely due to incorrect \textit{Gaia} periods (\citealt{Jayasinghe2022} found for $\sim11$\% of their detached eclipsing binary sample that the orbital periods from \textit{Gaia} SB1 disagreed with values from ASAS-SN). To verify this, we used orbital score values from \citet{Bashi2022} assigned to each \textit{Gaia} SB1, where the score (ranging from 0 to 1) corresponds to the validity of the orbital solution. They recommend a "clean score limit" of $0.587$, which yields a sensitivity of 80\% (that is, a sample of SB1's with scores $>0.587$ should have fewer than 20\% false orbits). 
Figure \ref{fig:sbbinary_rotation_period} separates each group by whether their score is $>0.587$. For most groups, the systems with higher scores lie overwhelmingly on one of the horizontal lines, while the systems with lower scores account for a majority of the scatter. Additionally, we assume that systems with reliable \textit{Gaia} periods also have reliable \textit{Gaia} eccentricities, and expect the spectroscopic binaries in our sample to be in close circular orbits after tidal synchronization. Most of the non-synchronous systems are also reported to be eccentric ($e>0.1$).

MS1 tends to have somewhat shorter $ P_{\rm{orb}}$ than $P_{\rm{rot}}$ and MS2 tends to be synchronized. 
G1 is strongly clustered on $P_{\rm{orb}}=2P_{\rm{rot}}$. While this is typical of ellipsoidal variables and contact binaries, our visual inspection of the light curves rules out misclassification of such stars. Instead, G1 likely consists of rotators in synchronized binaries with two dominant and roughly symmetrically-placed starspots such that the frequency of their observed change in brightness is doubled compared to their actual rotation period. 
G2 is strongly clustered on $P_{\rm{orb}}=P_{\rm{rot}}$ and so consists of synchronized systems. 
G3 also consists largely of stars with $P_{\rm{orb}}=P_{\rm{rot}}$. G3 systems with low orbital scores and non-zero eccentricity contribute significant scatter, especially for $P_{\rm{orb}}>P_{\rm{rot}}$. 
The G4b systems with both reliable and unreliable periods are strongly clustered at $P_{\rm{orb}}<P_{\rm{rot}}$, which means that G4b consists of younger giants in the process of tidal synchronization with their companions ("subsynchronous binaries"). This is consistent with the high-\texttt{VSCATTER} population of G4 having the lowest median $v\sin i$ of all the high-\texttt{VSCATTER} rotators in Table \ref{tab:med_fspot}.

We also checked the number of stars with RUWE $\geq$ 1.2 and the number of \textit{Gaia} astrometric binaries within each group. High RUWE and astrometric binaries are fairly common in the main sequence groups, but, as shown in Table \ref{tab:frac_comparisons}, there is little difference in the fractions with RUWE $\geq$ 1.2 or in the fractions of flagged astrometric binaries between the rotators and their twins. High RUWE and astrometric binaries are uncommon for all giant groups and their twins, as is expected given that they are generally more distant than the main sequence sample. Overall, the presence of a widely orbiting companion or tertiary seems to be unimportant in creating rotational variables.

\section{Conclusions}\label{section4}
Based on these results, we hypothesize that we have seven distinct groups of rotational variables: MS1, MS2s, MS2b, G1/G3, G2, G4s and G4b. We summarize our major findings about each group below:
\begin{enumerate}
    \item MS1 consists of main sequence K-M dwarfs with typical masses of $0.6$ to $0.8M_\odot$ based on the PARSEC isochrones. They generally are not (detectable) binaries, which for APOGEE means that the can only be in binaries with periods $>10^3$~days (see \citealt{Mazzola2020}). That they generally lie close to the main sequence means that few can have similar mass companions of any period unless they are sufficiently separated to be a spatially resolved binary.  They rotate relatively slowly (median period of 18.12 days), although not quite as slowly as their APOGEE twins.  The majority are not very heavily spotted.
    \item MS2s also consists of main sequence stars but with masses extending from the bottom of the main sequence to $0.8 M_{\odot}$.  They are also not binaries up to the same caveats as for MS1. They have faster rotation periods (median period of 8.08 days) and this is reflected in the different $v \sin i$ distributions. They also have higher spot fractions. 
    \item The properties of MS1 and MS2s are sufficiently disjoint that we are reasonably confident they are distinct. Figure \ref{fig:mcquillan_gap} shows the distribution of MS1, MS2s, MS2b, and the \textit{Kepler} sample of rotational variables from \citet{McQuillan2014} in rotation period and $T_{\rm{eff}}$. In this figure, we have separated MS2s from MS2b photometrically, as described in Section \ref{section3}, to maximize the sample size. The \textit{Kepler} sample  notably bifurcates in rotation period at $T_{\rm{eff}}<4500$K, possibly due to a transient phase of rapid mass-dependent spin-down across the gap \citep{McQuillan2014}. MS1 and MS2s seem to follow the direction of this "McQuillan gap," and while the MS2s stars span a wide range of temperatures, MS1 stars lie only at higher temperatures where \citet{McQuillan2014} did not find a bimodality in the \textit{Kepler} period distribution.
    
    The origin of the restricted temperature or mass range of MS1 as compared to MS2 is not presently clear. One possibility is that it is simply an amplitude-dependent selection effect against cool, long-period rotators, but a completeness study of the ASAS-SN sample is a major undertaking beyond the scope of this work. That they appear to lie on an extension of the \citet{McQuillan2014} period gap suggests that one difference between MS1 and MS2s is age, but we have no direct means of testing this other than the standard gyrochronological assessment.
\begin{figure*}
    \centering
    \includegraphics[width=\textwidth]{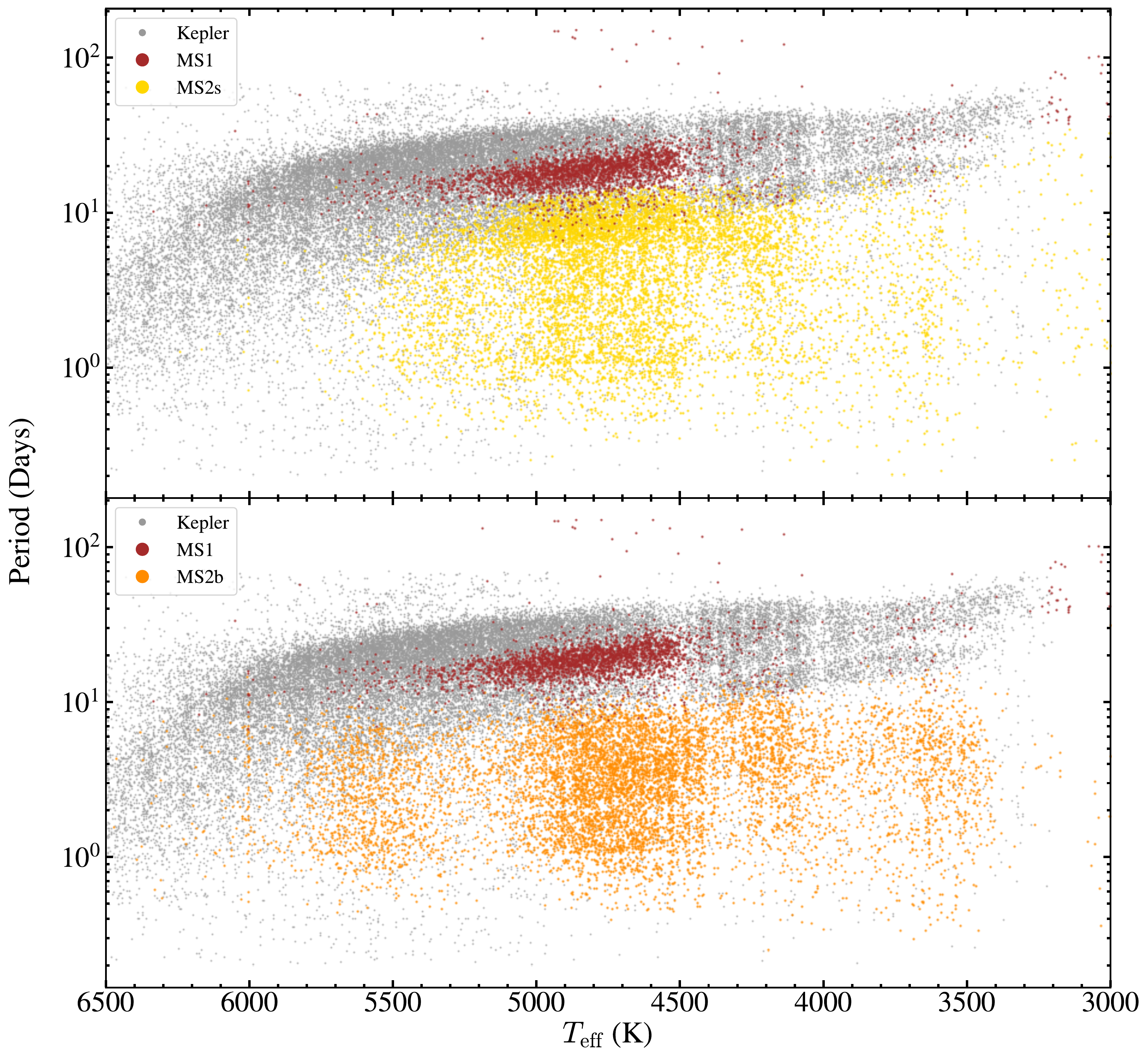}
    \caption{Distribution of the \textit{Kepler} rotational variables from \citet{McQuillan2014}, MS1, MS2s (top) and MS2b (bottom) in period and $T_{\rm{eff}}$.}
    \label{fig:mcquillan_gap}
\end{figure*}
    \item MS2b clearly is a different population than MS1 or MS2s. They are overwhelmingly synchronous binaries and many lie near the binary main sequence, indicating that the companion is of similar mass. Like the MS2s population, they seem to span the full dwarf mass range. They rotate a little faster than the MS2s stars, and have higher estimated spot filling fractions. In this case, however, the spectrum of the companion star may be contributing to the inferred spot fraction. That most of the MS2b stars have short periods suggests that the scatter of \textit{Kepler} stars to similar periods is also dominated by binaries, consistent with the findings in \citet{simonian2019}.
    \item  We suspect that G1 and G3 are really a single group of heavily spotted red giants in synchronized binaries (which we label G1/3 going forward). The G1 systems all have $P_{\rm{orb}}/P_{\rm{rot}}=2$ (see Figure \ref{fig:sbbinary_rotation_period}), which means that the true rotational period is twice that reported.  If we shift G1 in rotational period, they largely overlap with G3, and they have very similar luminosities, temperatures and gravities. The are not, however, truly identical after correcting the rotation period.  The G1 stars are more heavily spotted (median $f_{\rm{spot}}$ of 28\% versus 16\%) and have larger rotation velocities (median $v\sin i$ of 38 versus 21~km/s).  Since the rotational periods are similar, the higher $v \sin i$ suggests that the G1 stars are probably viewed more edge on. G1/3 stars are likely RS CVn stars \citep{hall1976}. \citet{leiner2022} noted that the rotation periods of RS CVn lie between 1-100 days and with magnitudes $-1 < M_G < 5$, both consistent with our sample of G1/3 stars. 
    \item  The G2 group likely consists of sub-subgiants (SSGs). \citet{leiner2022} found SSGs to have rotation periods overwhelmingly $<20$ days, consistent with our G2 sample, and with $3<M_G<5$. The G2 sample spans a somewhat broader range of $2<M_G<5$. \citet{leiner2022} also noted that their total RS CVn and SSG sample had increasing luminosity with rotation period, due to the necessity of a more massive companion to maintain tidal interaction in wider systems. In the left panel of Figure \ref{fig:perlum+cmd} we see a similar upward trend in the G1/3 and G2 groups. Finally, \citet{leiner2022} suggests a limit of $P\lesssim30$ days as a tidal circularization period for SSGs and the least luminous RS CVn, consistent with our boundary between the G2 and G4 groups (boundary \#3 in Table \ref{tab:cluster_lines}) at a period of $10^{1.5}\simeq30$~days.
    \item The G4s stars with low radial velocity scatter are likely recent merger products. 
    \item The G4b stars with high radial velocity scatter are sub-synchronous binaries (Figure \ref{fig:sbbinary_rotation_period}), beginning to tidally interact as they expand toward their companion. They are of intermediate luminosity (and so have intermediate evolution timescales) compared to G1/3 and G2, but have sufficiently wide orbits that their spin-up timescales are shorter than their evolutionary timescales. Neither the G4s recent mergers nor the G4b sub-synchronous binaries seem to have been previously recognized.
\end{enumerate}

There is enormous scope for expanding on this population study of rotational variables.  First, the ASAS-SN sample itself has considerable room for growth since the current sample is largely based on the older V-band ASAS-SN data and a small portion of the newer g-band data.  With the additional data it should not only be possible to expand the sample considerably but it should also be possible to push to lower amplitudes.  Rotational variables from brighter surveys like ASAS \citep{ASAS2002} or KELT \citep{kelt2018} could also be added, as well as the lower amplitude systems found by \textit{Kepler} (e.g. \citealt{McQuillan2014}) or TESS \citep{TESS2014}.  Except for the well-studied \textit{Kepler} field, there is less reason to expand to fainter stars because these will generally lack the ancillary spectroscopic data needed to study 
rotation rates, binarity, or composition.  The biggest immediate return is likely from including brighter systems, since this should significantly increase the numbers of systems with \textit{Gaia} spectroscopic orbits, the area of comparison where our samples are smallest.

Many of our conclusions rely primarily on the availability of APOGEE parameters, namely, \texttt{VSCATTER}, $v\sin i$, and $f_{\rm{spot}}$. APOGEE had specific targeting criterion which could bias our results, but we believe we cover any such biases both through confirming our observations in APOGEE data, if with lower precision, with information from \textit{Gaia}, and through comparison with the twin sample. APOGEE targeting was based on photometry \citep{Majewski2017}, so comparisons with the twins should be independent of APOGEE targeting choices. 

Additionally, APOGEE's spectroscopic data is significantly limited in availability compared to the size of our initial sample (APOGEE includes only $\sim4.4$\% of the ASAS-SN rotational variables from this work). However, the numbers of stars with ancillary spectroscopic data will rapidly increase with the SDSS Milky Way Mapper \citep{MWM2017} extension of the APOGEE survey \citep{Majewski2017} and \textit{Gaia} DR4.  The sources of the spectroscopic data could also be expanded to include surveys such as LAMOST  \citep{lamost2012} and DESI \citep{desimws2022, desi2022}.  The biggest impact will likely be \textit{Gaia} DR4 because it will provide a huge expansion of the numbers of systems with spectroscopic orbits and provide the actual RV measurements, which are needed to understand many of the rejected orbital solutions from DR3.  In particular, there are suggestions from Figure \ref{fig:sbbinary_rotation_period} that the G2 group might be slightly sub-synchronous, that the G3 group is a mixture of synchronized and unsynchronized systems, and that the sub-synchronous rotation of the G4b group is correlated with period. For now, we conservatively call the G2 and G3 groups synchronized, but with better orbital solutions we may find that
the situation is more complex.

We also did not explore the compositions of the variables both between the groups and with their twins.  The underlying reason is that APOGEE does not include stellar rotation in its models of the giants for computational reasons \citep{holtzman2018}.  This leads to biases on the inferred parameters because the broader lines created by the rapid rotation are interpreted as some other physics.  We see this here in the temperature and $\log g$ offsets between the giant groups and their twins. However, this problem also biases the abundances, and \citet{Patton2023} find that apparent abundance anomalies are a means of flagging rapidly rotating APOGEE giants.  Trying to understand how this problem would affect any elemental comparison seemed beyond the scope of this paper. This issue might also be one reason to include information from the large, lower resolution spectroscopic surveys like LAMOST and DESI, where the line broadening in giants due to rotation would be irrelevant.

\section*{Acknowledgements}

The authors thank Las Cumbres Observatory and its staff for their
continued support of ASAS-SN. CSK is supported by NSF grants AST-1814440 and
AST-1908570. Support for TJ was provided by NASA through the NASA Hubble Fellowship grant HF2-51509 awarded by the Space Telescope Science Institute, which is operated by the Association of Universities for Research in Astronomy, Inc., for NASA, under contract NAS5-26555. LC acknowledges support from TESS Cycle 5 GI program G05113 and NASA grant 80NSSC19K0597.

ASAS-SN is funded in part by the Gordon and Betty Moore
Foundation through grants GBMF5490 and GBMF10501 and by the
Alfred P. Sloan Foundation through grant G-2021-14192
to the
Ohio State University, the Mt. Cuba Astronomical Foundation,
the Center for Cosmology and AstroParticle Physics (CCAPP) at
OSU, the Chinese Academy of Sciences South America Center
for Astronomy (CAS-SACA), and the Villum Fonden (Denmark).
Development of ASAS-SN has been supported by NSF grant AST-
0908816, the Center for Cosmology and Astroparticle Physics,
Ohio State University, the Mt. Cuba Astronomical Foundation,
and by George Skestos.

\section*{Data Availability}
All data used in this study are public. 



\bibliographystyle{mnras}
\bibliography{ref}




\appendix




\bsp	
\label{lastpage}
\end{document}